\documentclass[]{JHEP}
\usepackage{epsfig}
\usepackage{cite}

\title{Hawking Radiation from a (4+n)-dimensional 
Black Hole: Exact Results
for the Schwarzschild Phase}

\author{C.M.~Harris$^a$ and P.~Kanti$^b$\\
$^a$Cavendish Laboratory, University of Cambridge, Madingley Road, 
Cambridge, CB3~0HE, UK.\\
$^b$Theory Division, CERN, 1211 Geneva 23, Switzerland.
}

\abstract{We start our analysis by deriving a {\it master equation} that
describes the motion of a field with arbitrary spin $s$ on a 3-brane embedded
in a non-rotating, uncharged $(4+n)$-dimensional black hole background. 
By numerical analysis, we derive exact results for the greybody factors
and emission rates for scalars, fermions and gauge bosons emitted directly
on the brane, for all energy regimes and for an arbitrary number $n$ of
extra dimensions. The relative emissivities on the brane for different types
of particles are computed and their dependence on the dimensionality of
spacetime is demonstrated --- we therefore conclude that both the amount and the
type of radiation emitted can be used for the determination of $n$ if the
Hawking radiation from these black holes is observed. The emission of
scalar modes in the bulk from the same black holes is also studied and 
the relative bulk-to-brane energy emissivity is accurately computed. We
demonstrate that this quantity varies considerably with $n$ but always
remains smaller than unity --- this provides firm support to earlier arguments
made by Emparan, Horowitz and Myers.}

 
\preprint{Cavendish-HEP-03/17 \\
		CERN-TH/2003-205}
		
\let\om=\omega	
\let\si=\sigma
\let\Ga=\Gamma	

\begin{document}

\section{Introduction}

Motivated by the desire to explain the hierarchy problem --- that is, the 
sixteen orders of magnitude difference between the electroweak energy scale and the
Planck scale --- models that postulate the existence of extra dimensions have
been revived and extensively studied during the last few years after
the work by Arkani-Hamed, Dimopoulos and Dvali (ADD) \cite{ADD} and
Randall and Sundrum (RS) \cite{RS} (for some early works, see \cite{early}).  
In the standard version of those works the Standard Model fields are
localized on a 3-brane, which plays the role of our 4-dimensional world,
while gravity can propagate both on the brane and in the bulk --- the
spacetime transverse to the brane. 
In theories with large extra dimensions the traditional Planck scale, 
$M_{Pl}\sim 10^{18}$ GeV, is only an effective energy scale derived
from the fundamental higher-dimensi\-onal one, $M_*$, through the following
relation \cite{ADD}
\begin{equation}
M_{Pl}^2 \sim  M_*^{n+2}\,R^n\,.
\label{MPl}
\end{equation}
The above relation involves the volume of the extra dimensions, $V \sim R^n$,
under the assumption that $R$ is the common size of all $n$ extra compact
dimensions. Therefore, if the volume of the internal space is large (i.e. if 
$R \gg \ell_{Pl}$, where $\ell_{Pl} =10^{-33}$ cm is the Planck length) 
then $M_*$ can be substantially lower than $M_{Pl}$.

In the regime $r\ll R$, the extra dimensions `open up' and gravity becomes
strong. Hence, Newton's law for the gravitational interactions in this
regime is modified, with the gravitational potential assuming a $1/r^{n+1}$
dependence on the radial separation between two massive particles. 
Experiments which measure the gravitational inverse-square law at short
scales can provide limits on the size of the extra dimensions, or equivalently
on the value of the fundamental scale $M_*$; for $n=2$ such measurements
give $M_*>~3.5$ TeV \cite{Hoyle} (the $n=1$ case has already been excluded
by astronomical data). Since gravitons can propagate both in the bulk and
on the brane, massive Kaluza-Klein (KK) graviton states can modify both
the cross sections of Standard Model particle interactions and astrophysical
or cosmological processes. The absence of signatures of production of either
real or virtual KK gravitons at colliders puts a relatively weak lower limit on
$M_*$ --- from 1.45~TeV (for $n=2$) to 0.6~TeV ($n=6$) \cite{colliders}.
Much more stringent constraints arise if one considers astrophysical or
cosmological processes; ignoring the systematic errors, these constraints
exclude by far even the $n=3$ case, while allowing models with 
$M_* \sim 1$~TeV and $n \geq 4$ \cite{astro}.

If extra dimensions with $R \gg \ell_{Pl}$  exist, then black holes with
a horizon radius $r_H$ smaller than the size of the extra dimensions $R$
are virtually higher-dimensional objects centered on the brane and extending
along the extra dimensions. It has been shown that these small black holes
have modified properties, i.e. they are larger and colder compared to a
4-dimensional black hole with exactly the same mass $M_{BH}$ \cite{ADMR}.
One striking consequence of the theories with large extra dimensions is that
the lowering of the fundamental gravity scale allows for the production of
such miniature black holes during high energetic scattering processes
with centre-of-mass energy $\sqrt{s} \gg M_*$ 
\cite{BF,GT,dl,voloshin,giddings,DE,GRW}. Arguments based
on Thorne's hoop conjecture \cite{Thorne} predict the creation of a black
hole in the case where any two partons of the colliding particles pass
within the horizon radius corresponding to their centre-of-mass energy.
The black holes created will have a mass equal to a fraction
of $\sqrt{s}$ varying from 0.84 down to 0.58 depending on the
details of the scattering process and the number of extra dimensions
\cite{fraction}. Provided that the mass of the black hole is larger than 
a few times the fundamental Planck mass, these objects could still be treated
semi-classically. Miniature black holes may be created in the atmosphere
of the earth (during scattering processes of cosmic rays) and at future
particle colliders, if the fundamental energy scale is low enough. Several
aspects of these processes have been studied in Refs. 
\cite{cosmic1,cosmic2,bhphen1,bhphen2,bhphen3,bhphen4}.

Once produced, these miniature black holes are expected to decay almost
instantaneously (typical lifetimes are $\sim 10^{-26}$~s). According to
Refs. \cite{GT,giddings}, the produced black holes will go through a number
of phases before completely evaporating. These are:

\begin{itemize}

\item
The {\it balding phase}\,: the black hole emits mainly gravitational radiation
and sheds the `hair' inherited from the original particles, and the asymmetry
due to the violent production process.

\item
The {\it spin-down phase}\,: the typically non-zero impact parameter
of the colliding partons leads to black holes with some angular momentum
about an axis perpendicular to the plane. During this phase, the black
hole loses its angular momentum through the emission of Hawking radiation
\cite{hawking} and, possibly, through superradiance.

\item
The {\it Schwarzschild phase}\,: a spherically-symmetric black hole loses
energy due to the emission of Hawking radiation. This results in the gradual
decrease of its mass and the increase of its temperature.

\item
The {\it Planck phase}\,: the mass and/or the Hawking temperature approach
the Planck scale --- a theory of quantum gravity is necessary to study this
phase in detail.

\end{itemize}

As in the 4-dimensional case \cite{page}, it is reasonable to expect
that the Schwarz\-schild
phase in the life of a small higher-dimensional black hole will be the
longer one, and will account for the greatest proportion of the mass loss 
through the emission of Hawking radiation. Throughout this paper,
we will focus on the emission of energy, in both brane-localized and bulk
modes, from a non-rotating, uncharged $(4+n)$-dimensional black hole.
Initially we will consider the emission of particle modes on the
brane, which is the most phenomenologically interesting effect since it
involves Standard Model particles. We will start by deriving a {\it master
equation} describing the motion of a field with arbitrary spin $s$ in the
spherically-symmetric black hole background induced on the brane --- this
will help resolve ambiguities present in similar equations which have 
previously appeared in the literature. We will then produce exact numerical 
results for the greybody factors and emission rates for each type of particle 
and for an arbitrary number $n$ of extra dimensions, and compare them with
those derived from earlier analytic studies of the
Schwarzschild phase \cite{kmr1, kmr2}. We will show
that the full analytic results derived in \cite{kmr2} closely follow
the exact results in the low- and intermediate-energy regimes while the
power expansion employed in \cite{kmr1} can accurately describe only a
very limited low-energy regime, especially in the case of scalar fields.
The major aim of this study is to demonstrate the dependence
of the above quantities on the dimensionality of spacetime in all energy
regimes.  This might serve as a tool for `reading' the number of extra
dimensions existing in nature if the spectrum of Hawking radiation
emitted by these small black holes is detected. In the framework of the
same analysis, we will also derive the relative emissivities for scalars,
fermions and gauge bosons and demonstrate how these change as the number
of extra dimensions projected onto the brane varies --- a potentially
additional signature to be used in the determination of $n$.

In Ref. \cite{emparan}, it was argued that the majority of energy during
the emission of Hawking radiation from a higher-dimensional black hole is
emitted into modes on the brane (i.e. Standard Model fermions and gauge bosons,
zero-mode gravitons and scalar fields). This argument was based on their
result that a single zero-mode brane particle carries as much energy as the
whole Kaluza-Klein (KK) tower of massive excitations propagating in the bulk
--- the abundance of brane modes compared to bulk modes then leads to the 
aforementioned conclusion. In this paper, we also study the emission of
bulk scalar modes from a $(4+n)$-dimensional black hole, and compare the
total bulk and brane emissivities in an attempt to confirm or disprove the
above argument. Our analysis improves the heuristic arguments
made in \cite{emparan} and for the first time provides exact and detailed
results concerning the relative bulk and brane emissivities in all energy
regimes and for various values of $n$. 

We start our analysis in section 2 by presenting the basic formulae and
assumptions for the gravitational background and emission rates from
small higher-dimensional black holes. In section 3, we present a
{\it master equation} describing the motion of scalars, fermions and
gauge bosons in the 4-dimensional induced background 
--- the basic steps of this calculation are given in the Appendix; in the
same section, 
we also define the corresponding greybody factor and emission rates and
discuss analytic and numerical methods for computing these quantities.
Exact numerical results for emission on the brane (for scalars, fermions and 
gauge bosons) are presented in section 4, as well as a discussion of the
relative emissivities and their dependence on $n$.  The
emission of bulk scalar modes is thoroughly studied in section 5, and our
results for the relative bulk-to-brane emissivities are also presented
in this section. Our conclusions are summarized in section 6.

\section{Basic Formulae and Assumptions}
\label{basic}

Let us start with the form of the gravitational background around a
non-rotating, uncharged $(4+n)$-dimensional Schwarzschild black hole. The
line-element is given by \cite{myers}
%%%%%%%%%%%
\begin{equation}
ds^2=- h(r)\,dt^2 + h(r)^{-1}\,dr^2 + r^2\,d \Omega^2_{2+n}\,,
\label{metric-D}
\end{equation}
%%%%%%%%%%%
where 
\begin{equation}
h(r) = 1-\biggl(\frac{r_H}{r}\biggr)^{n+1}\,,
\label{h-fun}
\end{equation}
%%%%%%%%%%
and with the angular part given by
%%%%%%%%
\begin{eqnarray} \hspace*{-0.3cm}
d\Omega_{2+n}^2=d\theta^2_{n+1} + \sin^2\theta_{n+1} \,\biggl(d\theta_n^2 +
\sin^2\theta_n\,\Bigl(\,... + \sin^2\theta_2\,(d\theta_1^2 + \sin^2 \theta_1
\,d\varphi^2)\,...\,\Bigr)\biggr)\,.
\end{eqnarray}
%%%%%%%%%%%
In the above, $0 <\varphi < 2 \pi$ and $0< \theta_i < \pi$, for 
$i=1, ..., n+1$. By using an analogous approach to the usual 4-dimensional
Schwarzschild calculation, i.e. by applying Gauss' law in the 
$(4+n)$-dimensional spacetime, we obtain the following horizon radius 
%%%%%%%%%%%%
\begin{equation}
r_H= \frac{1}{\sqrt{\pi}M_*}\left(\frac{M_{BH}}{M_*}\right)^
{\frac{1}{n+1}}\left(\frac{8\Gamma\left(\frac{n+3}{2}\right)}{n+2}\right)
^{\frac{1}{n+1}},
\end{equation}
where $M_{BH}$ is the mass of the black hole. The black holes being 
considered in this work are assumed to have horizon radii satisfying
the relation $\ell_{Pl} \ll r_H \ll R$.
The former inequality guarantees that no quantum corrections are important
in our calculations, while the latter is necessary for these black holes to
be considered as higher-dimensional objects. The tension of the brane
on which the black hole is centered is assumed to be much smaller than the
black hole mass and thus it can be ignored in our analysis.

A black hole of a particular horizon radius $r_H$ is characterized by a
Hawking temperature related by the expression
%%%%%%%%%%
\begin{equation}
T_{BH}=\frac{(n+1)}{4\pi\,r_H}\,.
\end{equation}
%%%%%%%%%%%
The above temperature gives rise to {\it almost} blackbody radiation.
The flux spectrum, i.e. the number of particles emitted per unit time,
is given by \cite{hawking}
%%%%%%%%%%%%%
\begin{equation}
\label{flux}
 \frac{dN^{(s)}(\om)}{dt} = \sum_{\ell} \sigma^{(s)}_{\ell}(\om)\,
{1  \over \exp\left(\om/T_{BH}\right) \pm 1} 
\,\frac{d^{n+3}k}{(2\pi)^{n+3}}\,,
\end{equation}
%%%%%%%%%%%%%%
while the power spectrum, i.e. the energy emitted per unit time, is
%%%%%%%%%%%%%
\begin{equation}
\frac{dE^{(s)}(\om)}{dt} = \sum_{\ell} \sigma^{(s)}_{\ell}(\om)\,
{\om  \over \exp\left(\om/T_{BH}\right) \pm 1}\,\frac{d^{n+3}k}{(2\pi)^{n+3}}\,.
\label{power}
\end{equation}
%%%%%%%%%%%%%
In the above, $s$ is the spin of the degree of freedom being considered
and $\ell$ is the angular momentum quantum number. The spin statistics factor
in the denominator is $-1$ for bosons and $+1$ for fermions. For 
massless particles, $|k|=\om$ and the phase-space integral reduces to
an integral over $\omega$. The term in
front, $\sigma^{(s)}_{\ell} (\om) $,  is the so-called `greybody' factor
\footnote{The quantity $\sigma^{(s)}_{\ell} (\om) $ is alternatively called 
the absorption cross section. It is also common in the literature to
refer to the absorption probability $|{\cal A}^{(s)}_\ell|^2$, related to
$\sigma^{(s)}_{\ell} (\om) $ through Eq. (\ref{greybody}), as the greybody
factor.} which encodes valuable information for the structure of the spacetime 
around the black hole (which emits the Hawking radiation) including the 
dimensionality of spacetime. This quantity can be determined by solving the 
equation of motion of a particular degree of freedom in the aforementioned 
background and computing the corresponding absorption coefficient
${\cal A}^{(s)}_\ell$. Then, we may write \cite{GTK} 
%%%%%%%%%%%%%
\begin{equation}
\si^{(s)}_\ell(\om) = \frac{2^{n}\pi^{(n+1)/2}\,\Ga[(n+1)/2]}{ n!\, \om^{n+2}}\,
\frac{(2\ell+n+1)\,(\ell+n)!}{\ell !}\, |{\cal A}^{(s)}_\ell|^2\,.
\label{greybody}
\end{equation}
%%%%%%%%%%% 
As the decay progresses, the black hole mass decreases and the Hawking temperature rises. It is usually assumed that a quasi-stationary approach 
to the decay is valid --- that is, the black hole has time to come into 
equilibrium at each new temperature before the next particle is emitted. We 
will make this assumption also here.

We would like finally to stress that Eqs. (\ref{flux}) and (\ref{power}) 
refer to individual degrees of freedom and not to elementary
particles, like electrons or quarks, which contain more than one polarization.
Combining the necessary degrees of freedom and their corresponding flux
or power spectra, the number of elementary particles produced, and the
energy they carry, can be easily computed.  For more information on this,
we refer the reader to Ref. \cite{HRW} where a Black Hole Event Generator
has been constructed.  This simulates both the production and decay
of small black holes at hadronic colliders and, by using the results
for the greybody factors derived here, provides estimates for the
number and spectra of the different types of elementary particles produced.

%%%%%%%%%%%%%%%%%%%%%%%%%%%%%%%%%%%%%%%%%%%%%%%%%%%%%%%%%%%%%%%%%%%%%%%%%%%%%%%%

\section{Greybody factors for Emission on the Brane}

The greybody factors modify the spectrum of emitted particles from that of a
perfect thermal blackbody \emph{even} in four dimensions \cite{hawking}. 
For a 4-dimensional Schwarzschild black hole, geometric arguments show
that, in the high-energy regime, %with $\omega\gg T_H$, 
$\Sigma_{l}\,|{\cal A}_\ell|^2 \propto(\omega r_H)^2$ which means that
the greybody factor at high energies is independent of $\omega$ and the
spectrum is exactly like that of a blackbody for every particle species
\cite{MTW, sanchez, page}. The low-energy behaviour, on the other hand, is
strongly spin-dependent. A common feature for fields with spin $s=0$,
$\frac{1}{2}$ and 1 is that the greybody factors reduce
the low-energy emission rate significantly below the geometrical optics value 
\cite{page, macG}.  The result is that both the power and flux spectra peak at 
higher energies than those for a blackbody at the same temperature.  The spin
dependence of the greybody factors means that they are necessary to determine 
the relative emissivities of different particle types from a black hole (at
present these are only available in the literature for the 4D case
\cite{page,sanchez}).

The procedure for calculating greybody factors needs to be generalised to
include emission from small higher-dimensional black holes with $r_H < R$
which emit radiation either in the bulk or on the brane. Analytical studies
\cite{kmr1, kmr2} have shown that the greybody factors in that case have a
strong dependence on the number of extra dimensions. The dependence
on both energy and the number of dimensions means that the exact form of the
greybody factors should be taken into account in any attempt to determine the
number of extra dimensions by studying the energy spectrum of particles emitted
from a higher-dimensional black hole. In this section, as well as in section 4,
we will focus on the emission of brane-localised modes leaving the
study of bulk emission and of relative bulk-to-brane emissivity for section 5.

The brane-localised modes propagate in a 4-dimensional black hole background
which is the projection of the higher-dimensional one, given in Eq. (\ref{metric-D}), onto the brane. The induced metric tensor follows by fixing
the values of the extra angular coordinates, $\theta_i=\pi/2$ for $i \geq 2$,
and it is found to have the form
%%%%
\begin{equation}
ds^2=-h(r)\,dt^2+h(r)^{-1}dr^2+r^2\,(d\theta^2 + \sin^2\theta\,d \varphi^2)\,.
\label{non-rot}
\end{equation}

The greybody factors are determined from the amplitudes of in-going and
out-going waves at infinity so the essential requirement is to solve the
equation of motion of a particle propagating in the above background. 
For this purpose, we have derived in the Appendix a generalised 
{\it master equation} for a particle with
arbitrary spin $s$, similar to the one derived by Teukolsky \cite{Teukolsky}.
For $s=\frac{1}{2}$ and 1, the equation of motion has been
derived by using the Newman-Penrose method \cite{np, Chandra}, while for $s=0$
the corresponding equation follows by the evaluation of the double covariant
derivative $g^{\mu\nu} D_\mu D_\nu$ acting on the scalar field. The derived
{\it master equation} is separable in each case and, by using the factorization
%%%%%%%%%%%%%%
\begin{equation}
\Psi_s=e^{-i\omega t}\,e^{im\varphi}\,R_s(r)\,S_{s,l}^m(\theta)\,,
\end{equation}
%%%%%%%%%%%%
\noindent we obtain the radial equation
%%%%%%%%%%%
\begin{equation}
\label{radial}
\Delta^{-s} \frac{d}{dr}\left(\Delta^{s+1}\,\frac{d R_s}{dr}\right)+
\left(\frac{\omega^2 r^2}{h}+2i\omega s r-\frac{is\omega r^2 h'}{h}+
s(\Delta''-2)-\lambda_{sl} \right)R_s (r)=0,
\end{equation}
where $\Delta=hr^2$.  The corresponding angular equation has the form
%%%%%%%%%%%
\begin{equation}
\frac{1}{\sin\theta}\,\frac{d}{d\theta}\left(\sin\theta\,\,
\frac{dS_{s,l}^m}{d\theta}\right)+
\left(-\frac{2ms\cot\theta}{\sin\theta}-\frac{m^2}{\sin^2\theta}+
s-s^2\cot^2\theta+\lambda_{sl}\right)S_{s,l}^m(\theta)=0\,,
\label{angular}
\end{equation}
%%%%%%%%%%%%%%%%
where $Y_{s,l}^m=e^{im\varphi}\,S_{s,l}^m(\theta)$ are known as the
spin-weighted spherical harmonics and $\lambda_{sl}$ is a separation constant
which is found to have the value $\lambda_{sl}=l(l+1)-s(s+1)$ \cite{goldberg}.
For $s=0$, Eq. (\ref{radial}) reduces as expected to Eq. (41) of 
Ref. \cite{kmr1} which was used for the analytical study of the emission of 
brane-localised scalar modes from a spherically-symmetric higher-dimensional
black hole.  Under the redefinition $R_s=\Delta^{-s} P_s$, Eq. (\ref{radial})
assumes a form similar to that of Eq. (11) of Ref. \cite{kmr2} which was
used for the study of brane-localised fermion and gauge boson emission.
The two equations differ due to an extra term in the expression of the latter
one, which although vanishing for $s=\frac{1}{2}$ and 1 (thus leading to the
correct results for the greybody factors for fermion and gauge boson fields)
gives a non-vanishing contribution
for all other values of $s$. Therefore, the generalized equation derived by
Cvetic and Larsen \cite{CL} can not be considered as a {\it master equation}
valid for all types of fields. A similar equation was derived in 
Ref. \cite{iop} but due to a typographical error 
the multiplicative factor $s$ in front of the $\Delta''$-term is missing,
thus leading to an apparently different radial equation for all values of
$s$. Therefore, before addressing the question of the derivation of the
exact forms of the greybody factors in the brane background, the derivation of
a consistent {\it master equation} was imperative. This task was indeed
performed, with some of the steps of the calculation presented in the Appendix,
and has led to the aforementioned Eqs. (\ref{radial}) and (\ref{angular}).

For the derivation of the greybody factors associated with the emission of
fields from the projected black hole, we need to know the asymptotic solutions
of (\ref{radial}) both as $r\rightarrow r_H$ and as $r\rightarrow \infty$. 
In the former case, the solution is of the form
%%%%%%%%%
\begin{equation}
\label{near}
R_s^{(h)}=A_{in}^{(h)}\,\Delta^{-s}\,e^{-i\omega r^{*}}+
A_{out }^{(h)}\,e^{i\omega r^{*}},
\end{equation}
%%%%%%%%%%
\noindent where
%%%%%%%%%%%%
\begin{equation}
\label{rstarr}
\frac{dr^*}{dr}=\frac{1}{h(r)}\,.
\end{equation}
%%%%%%%%%
We impose the boundary condition that there is no out-going solution near the
horizon of the black hole, and therefore set $A_{out}^{(h)}=0$.
The solution at infinity is of the form
%%%%%%%%%%%
\begin{equation}
\label{far}
R_s^{(\infty)}=A_{in}^{(\infty)}\,\frac{e^{-i\omega r}}{r}+
A_{out}^{(\infty)}\,\frac{e^{i\omega r}}{r^{2s+1}}\,,
\end{equation}
and comprises both in-going and out-going modes. 

The greybody factor $\sigma_{\ell}(\om)$ for the emission of brane-localised
modes is related to the energy absorption coefficient ${\cal A}_\ell$
through the simplified relation
%%%%%%%%
\begin{equation}
\hat \si^{(s)}_\ell(\om) =\frac{\pi}{\om^2}\,(2 \ell +1)\,
|\hat {\cal A}^{(s)}_\ell|^2\,,
\label{brane-loc}
\end{equation}
%%%%%%%%%%%%
where henceforth quantities with a `hat' will denote quantities associated with
the emission of brane-localised modes. The above relation follows from
Eq. (\ref{greybody}) by setting $n=0$ since the emission of brane-localised
modes is a 4-dimensional process. In the same way, the power spectrum 
of the Hawking radiation emitted on the brane can be computed by the
4-dimensional expression 
%%%%%%%%%%%%%
\begin{equation}
\frac{d \hat E^{(s)}(\om)}{dt} = \sum_{\ell} \hat \sigma^{(s)}_{\ell}(\om)\,
{\om  \over \exp\left(\om/T_{BH}\right) \pm 1}\,\frac{d^3 k}{(2\pi)^{3}}\,.
\label{decay-brane}
\end{equation}
%%%%%%%%%%%%% 
Note, however, that the absorption
coefficient $\hat {\cal A}_\ell$ still depends on the number of extra
dimensions since the projected metric tensor (\ref{non-rot}) carries a
signature of the dimensionality of the bulk spacetime through the expression
of the metric function $h(r)$. The absorption coefficient itself is
defined as
%%%%%%%%%%%
\begin{equation}
|\hat {\cal A}^{(s)}_\ell|^2=1-\frac{{\cal F}_{out}^{(\infty)}}
{{\cal F}_{in}^{(\infty)}}=
\frac{{\cal F}_{in}^{(h)}}{{\cal F}_{in}^{(\infty)}}\,,
\label{absorption}
\end{equation}
in terms of the energy fluxes evaluated either at infinity or at the horizon.
The two definitions are related by simple energy conservation and lead to
the same results. Both of them may be used for the determination of the
absorption coefficient, depending on the type of particle emitted and the
method of the study, numerical or analytic, which is followed.

\subsection{Analytic calculation}

The greybody factors have been determined analytically for the 4-dimensional
case in Refs. \cite{page, sanchez} both for a rotating and non-rotating black
hole. In the ($4+n$)-dimensional case, Refs. \cite{kmr1,kmr2} have provided
analytic expressions for the greybody factors for the emission of scalars,
fermions and gauge bosons from a higher-dimensional Schwarzschild-like
black hole emitting radiation both in the bulk and on the brane, and for 
an arbitrary number of extra dimensions $n$. Ref. \cite{iop} presented
analytic results for the
greybody factors for the emission of brane-localized modes from a Kerr-like
black hole in the particular case of $n=1$. All the above results were derived
in the low-energy approximation where the procedure used is as follows:

\begin{itemize}
\item{Find an analytic solution in the near-horizon regime and expand as
in-going and out-going waves so that the $A^{(h)}$ coefficients can be extracted.}
\item{Apply the boundary condition on the horizon so that the wave is purely
out-going.}
\item{Find an analytic solution in the far-field regime and again expand as
in-going and out-going waves.}
\item{Match the two solutions in the intermediate regime.}
\item{Extract ${\cal A}_\ell$ and expand in powers of $(\omega r_H)$.}
\end{itemize}

The solution obtained by following the above approximate method is a power
series in $\omega r_H$, which is only valid for low energies and expected to
significantly deviate from the exact result as the energy of the emitted
particle increases (this was pointed out in \cite{page,macG} for the
4-dimensional case). In \cite{kmr2}, the full analytic result for the absorption
coefficient (before the final expansion was made) was used for the evaluation
of the emission rates for all particle species. The range of validity of these
results, although improved compared to the power series,  was still limited since the assumption of small
$\omega r_H$ was {\it still} made during the matching of the two solutions
in the intermediate regime. As the equations of motion for all types of particles
are too cumbersome to solve analytically for any value of $\om r_H$, it becomes
clear that only an exact numerical analysis can yield the full results for 
greybody factors and emission rates. 

%%%%%%%%%%%%%%%%%%%%%%%%%%%%%%%%%%%%%%%%%%%%%%%%%%%%%%%%%%%%5

\subsection{Numerical calculation}

There are various numerical issues which arise while trying to do the full
calculation of the absorption coefficient; the complexity of these problems
strongly depends on the spin $s$ of the emitted particle.  The usual numerical 
procedure starts by applying the required horizon boundary condition (a 
vanishing out-going wave) to $R_s (r)$.  Then, the solution is integrated out 
to `infinity' and the asymptotic coefficient $A_{in}^{(\infty)}$ is extracted 
in terms of which ${\cal A}_\ell$ can be calculated.

By looking at the asymptotic solution at infinity, Eq. (\ref{far}), one can
see that, for the scalar case, the in-going and out-going waves are of
comparable magnitude.  Therefore it is relatively easy to extract the
coefficients at infinity and thus determine the greybody factor.  However for 
fields with non-vanishing spin, i.e. fermions, vector bosons and
gravitons, this is not an easy task. First of all, different components carry
a different
part of the emitted field: the upper component $\Psi_{+s}$ consists mainly of
the in-going wave with the out-going one being greatly suppressed, while for
the lower component $\Psi_{-s}$ the situation is reversed (for a field with
spin $s \neq 0$, only the upper and lower components are radiative). 
Distinguishing between the two parts of the solution, in-going
and out-going, is not an easy task no matter which component we look at, and it
becomes more difficult as the magnitude of the spin increases.  In addition,
the choice of either positive or negative $s$ to extract the greybody factor,
i.e. using either the upper or the lower component, affects
the numerical issues.

If $s$ is negative, then the horizon boundary condition is easy to apply 
because components of the in-going solution on the horizon will be 
exponentially suppressed [see Eq. (\ref{near}) above].  However, negative $s$ 
also means that at infinity the out-going solution is enhanced by $r^{-2s}$ 
with respect to the in-going one [see Eq. (\ref{far})], which makes accurate 
determination of $A_{in}^{(\infty)}$ very difficult.  Hence the negative $s$ 
approach is not used in this work.

On the other hand if $s$ is positive, it is easy to accurately extract 
$A_{in}^{(\infty)}$ because the out-going solution at infinity is suppressed by
a factor of $r^{-2s}$.  However, close to the horizon the out-going solution is
exponentially smaller than the in-going one. 
This means that the solution for $R_s(r)$ can be easily contaminated by 
components of the out-going solution. This problem becomes worse for larger 
$s$ and first becomes significant for $s=1$.

For $n=0$, various methods (see \emph{e.g.}\ \cite{tp3}) have been used to 
solve the numerical problems which arise in the gauge boson case.  The 
approach of Bardeen in \cite{tp3} is also applicable for $n=1$, but no 
analogous method was found for higher numbers of extra dimensions. Here, an 
alternative transformation of the radial equation was used instead.
Writing $y=r/r_H$ and $R_1=y\,F(y)\,e^{-i\omega r^*}$, the wave equation becomes
%%%%%%%%%%%%
\begin{equation}
(hy^2)\,\frac{d^2F}{dy^2}+2y\,(h-i\omega r_H y)\,\frac{dF}{dy}-l(l+1)F=0.
\end{equation}
%%%%%%%%%%%%
Since on the horizon $y=1$ and $h=0$, the boundary conditions become $F(1)=1$ 
and
%%%%%%%%%%%%
\begin{equation}
\left. \frac{dF}{dy} \right|_{y=1}=\frac{i l(l+1)}{2 \omega r_H}.
\end{equation}
%%%%%%%%%%%%%%

For fermions, no such transformation was necessary and so the radial
equation of Eq. (\ref{radial}) was used.  However the application of the 
boundary condition at the horizon is made slightly easier by the 
transformation $P_s=\Delta^s R_s$, which means that the asymptotic solution at
the horizon becomes
%%%%%%%%%%%
\begin{equation}
\label{pnear}
P_s^{(h)}=A_{in}^{(h)}\,e^{-i\omega r^{*}}+
A_{out }^{(h)}\,\Delta^s \,e^{i\omega r^{*}}.
\label{nh-new}
\end{equation}
%%%%%%%%%%%%
Since we require $A_{out}^{(h)}=0$, suitable boundary conditions to apply, when
solving the differential equation for $P_s$, are that, as $r \rightarrow r_H$,
%%%%%%%%%%
\begin{equation}
P_s=1\,,
\end{equation}
%%%%%%%%%%%%%%
while using Eq. (\ref{rstarr}), we also obtain
%%%%%%%%%%%%%%%
\begin{equation}
\frac{dP_s}{dr}=-i\omega\, \frac{dr*}{dr}=-\frac{i\omega}{h(r)}\,.
\end{equation}
%%%%%%%%%%%%
The above boundary conditions ensure that $|A_{in}^{(h)}|^2=1$. 
The asymptotic form for $P_s$ at infinity now looks like
%%%%%%%%%%%%
\begin{equation}
\label{pfar}
P_s^{(\infty)}=A_{in}^{(\infty)}\,\frac{e^{-i\omega r}}{r^{1-2s}}+
A_{out}^{(\infty)}\, \frac{e^{i\omega r}}{r}\, 
\label{ff-new}
\end{equation}
%%%%%%%%%%%
since $\Delta \rightarrow r^2$ as $r \rightarrow \infty$. 

There are various considerations which must be taken into account in order to
obtain results to the required accuracy (at least three significant figures).
Firstly, although the horizon boundary condition can not be applied exactly at
$r_H$ (due to singularities in the boundary condition and the differential 
equation) the error introduced by applying the condition at $r=r_c$ (where $r_c-r_H \ll r_H$) must be small.  This can be checked by studying changes in the
greybody factors for order of magnitude changes in $r_c-r_H$.  Similarly it 
must be checked that the value of $r$ used as an approximation for `infinity' 
does not introduce errors which will affect the accuracy of the result.  Care 
must also be taken that the numerical integration procedure is sufficiently 
accurate out to large values of $r$ so that significant integration errors are
avoided.  Finally, for each energy being considered, enough angular momentum 
modes must be included in the summation so that only higher modes which do not
contribute significantly are neglected.  For the higher values of $\omega r_H$
considered in this work this means that the contributions from in excess of 
ten angular momentum modes are required.

\section{Numerical Results for Brane Emission}

In this section, we proceed to present our results for the greybody factors
and emission rates for brane-localised scalar, fermion and gauge boson
fields, as obtained by numerically solving the corresponding equations of 
motions.
The definition of the absorption coefficient ${\cal A}_\ell$ is different
for each type of field due to their different asymptotic behaviour in
the far-field regime and also due to more fundamental differences between 
fields with zero and non-zero spin. We will, therefore, consider each case separately:

\subsection{Spin 0 fields}

The numerical integration of Eq. (\ref{radial}) for $s=0$ yields the solution
for the radial function $R_0(r)$ which smoothly interpolates between the
asymptotic solutions (\ref{nh-new}) and (\ref{ff-new}) in the near-horizon
and far-field regimes respectively. The absorption coefficient is easily
defined in terms of the in-going and out-going energy fluxes at infinity,
or equivalently by the corresponding wave amplitudes in the same asymptotic
regime, given by $A_{in}^{(\infty)}$ and $A_{out}^{(\infty)}$ respectively.
We may thus write Eq. (\ref{absorption}) in the form \cite{kmr1}
%%%%%%%%%%%
\begin{equation}
|\hat {\cal A}^{(0)}_\ell|^2=1-|\hat {\cal R}^{(0)}_\ell|^2= 
1-\Biggl|\frac{A_{out}^{(\infty)}}{A_{in}^{(\infty)}}\Biggr|^2\,,
\label{scalars}
\end{equation}
%%%%%%%%%%
where $\hat {\cal R}_\ell$ is the corresponding reflection coefficient. 

\FIGURE[b]{
\scalebox{0.5}{\rotatebox{-90}{\includegraphics[width=\textwidth]{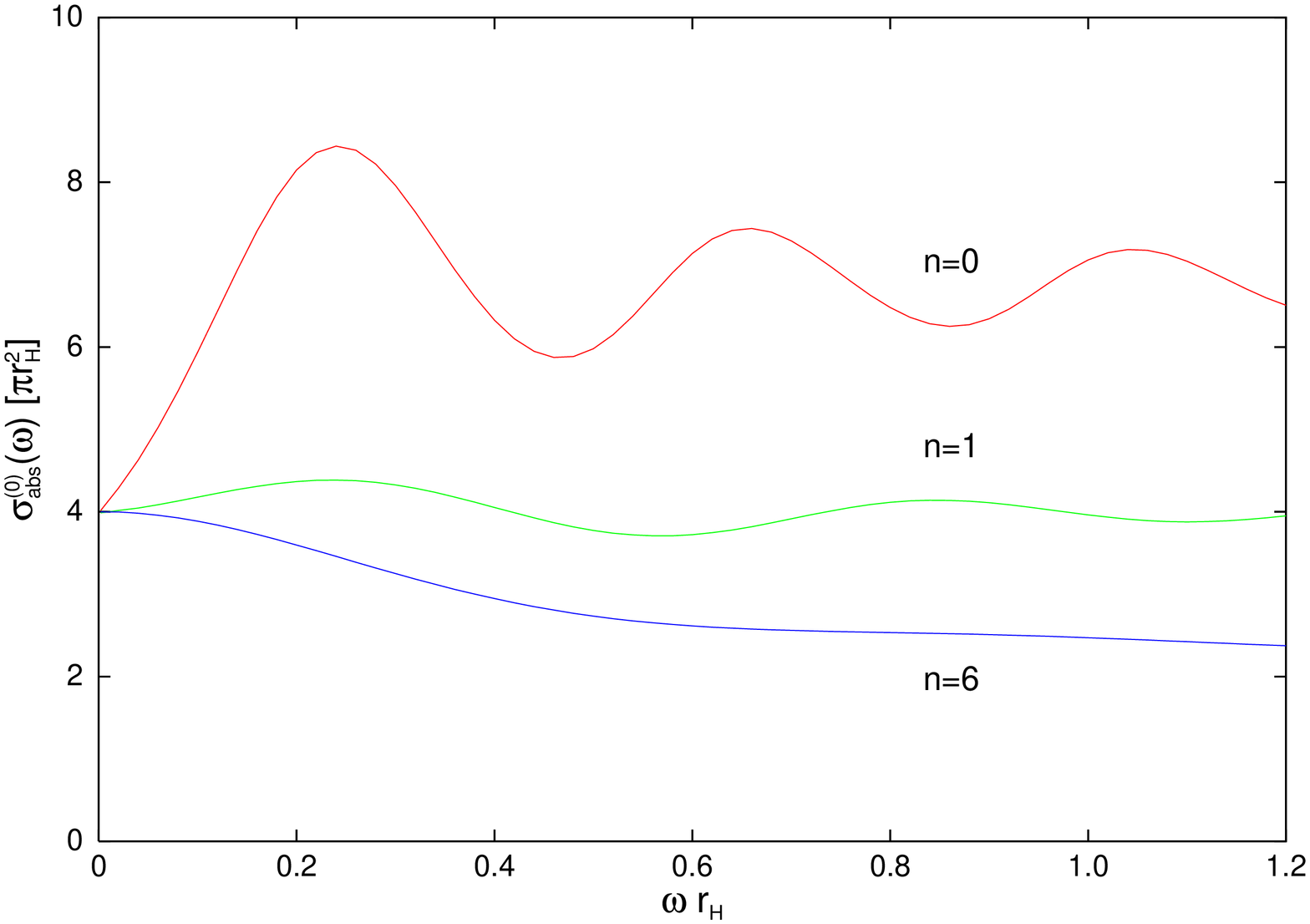}}}
\caption{Greybody factors for scalar emission on the brane from a $(4+n)$D
black hole.}
\label{grey0}}

The plot presented in Figure \ref{grey0} shows, for different values of $n$,
the greybody factors for the emission of scalar fields on the brane (note
that, throughout our numerical analyses, the horizon radius $r_H$ is an
arbitrary input parameter which remains fixed). The greybody
factor is derived by using Eq. (\ref{brane-loc}), which is valid for the
emission of brane-localised modes, and summing over the angular momentum
number $\ell$ (for completeness, the plots include values of $n$ ruled out
on astrophysical grounds, i.e. $n=1,2$). For $n=0$ and
$\omega r_H \rightarrow 0$, the greybody factor assumes a non-zero value 
which is equal to $4\pi r_H^2$ --- that is, the greybody factor for scalar 
fields with a very low energy is given exactly by the area of the black hole 
horizon. As the energy increases, the factor soon starts oscillating
around the geometrical optics
limit $\sigma_g = 27 \pi r_H^2/4$ which corresponds to the spectrum of a
black-body with an absorbing area of radius $r_c=3 \sqrt{3}\,r_H/2$
\cite{sanchez,MTW}. If extra dimensions are present, the greybody factor
starts from the same asymptotic low-energy value, for any value of $n$,
and it again starts oscillating around a limiting high-energy value, which
is always lower than the 4-dimensional one.
This is because the effective radius $r_c$ depends on the
dimensionality of the bulk spacetime through the metric tensor of the
projected spacetime (\ref{non-rot}) in which the particle moves. For
arbitrary $n$, it adopts the value \cite{emparan}
%%%%%%%%%%%
\begin{equation}
r_c=\biggl(\frac{n+3}{2}\biggr)^{1/n+1}\,\sqrt{\frac{n+3}{n+1}}\,\,r_H\,.
\label{effective}
\end{equation}
%%%%%%%%%%%%
The above quantity keeps decreasing as $n$ increases causing the
asymptotic greybody factor, $\sigma_g=\pi r_c^2$, to become more and more
suppressed as the number of extra dimensions projected onto the
brane gets larger. 

The power series expression of the greybody factor determined in \cite{kmr1}
matches the exact solution only in a very limited low-energy
regime. In the limit $\omega r_H \rightarrow 0$, the asymptotic
value $4\pi r_H^2$ is recovered as expected; however, as the energy increases
the exact solution rapidly deviates from the behaviour dictated by the
dominant term in the $\omega r_H$ expansion. This was first
demonstrated in \cite{kmr2}, where the full analytic result for the greybody
factor was determined. The behaviour depicted in Figure 3 of Ref. \cite{kmr2}
is much closer to the exact one, shown here in Figure \ref{grey0}, and
successfully reproduces the qualitative features including the suppression of
the greybody factor as the dimensionality of the bulk spacetime increases.
Nevertheless, as previously stated, even that result breaks down in the
high-energy regime leaving the exact numerical solution produced here
as the only reliable source of information concerning the form of the
greybody factor throughout the energy regime.

\FIGURE[b]{ \hspace{-0.8cm}
\scalebox{0.5}{\rotatebox{0}{\includegraphics[width=23cm, height=15cm]{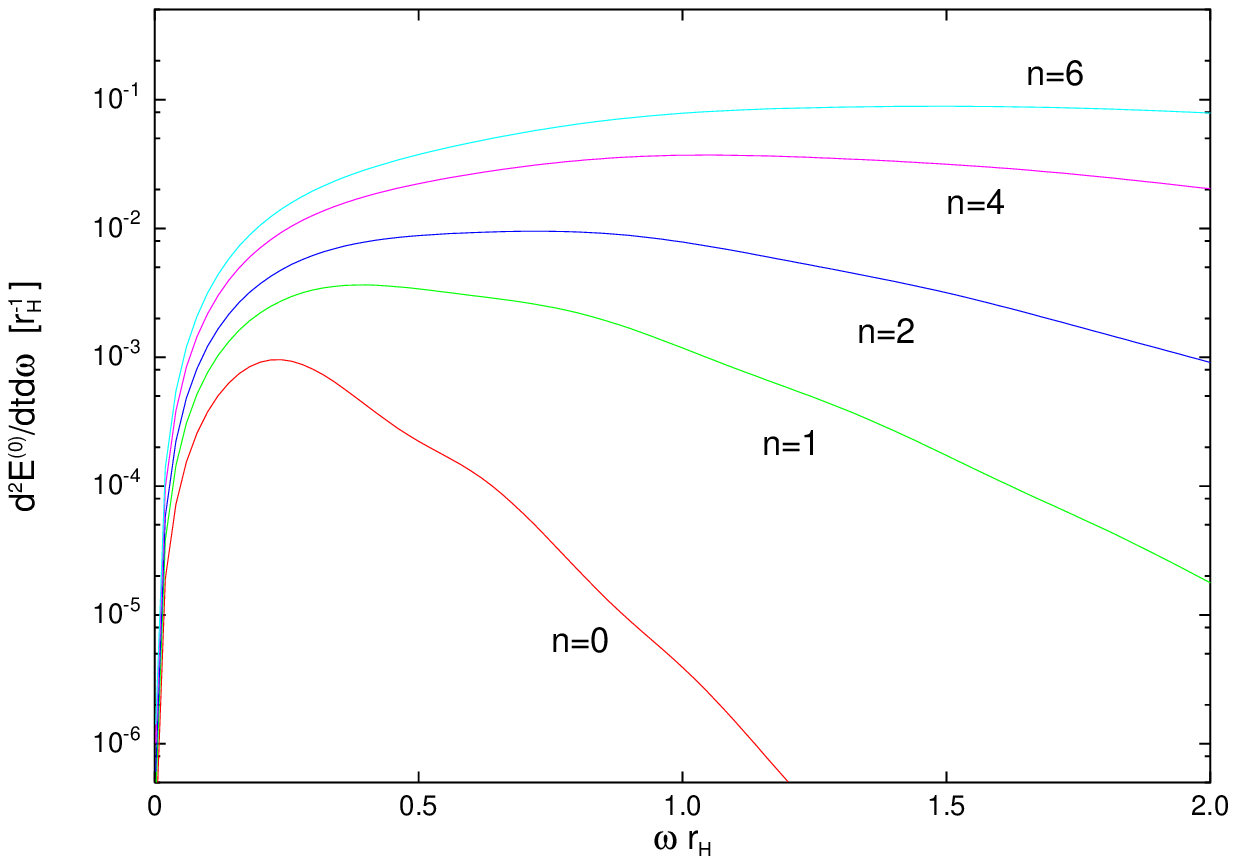}}}
\caption{Energy emission rates for scalar fields on the brane from
a $(4+n)$D black hole.}
\label{sc-dec}
}

Proceeding to compute the energy emission rate for scalar fields on the
brane, by using Eq. (\ref{decay-brane}), we find that the suppression
of the greybody factor with $n$ does not necessarily lead to the suppression
of the emission rate itself. The behaviour of the differential energy
emission rate in time unit $dt$ and energy interval  $d\omega$ is given in
Figure \ref{sc-dec}. The increase in the temperature of the black hole, and
thus in its emissivity power, as $n$ increases overcomes the decrease in
the value of the greybody factor and leads to a substantial enhancement of
the energy emission rate. As it becomes clear from Figure \ref{sc-dec}, the
increase in the number of dimensions projected onto the brane causes the
enhancement of the peak value of the emission curve by many orders of
magnitude, when compared to the 4-dimensional case. In addition, it leads
to a Wien's type of displacement of the peak, i.e. to the shift of the peak
towards higher values of the energy parameter $\omega r_H$, reflecting
the increase in the temperature of the radiating body.

In order to be able to clearly perceive the amount of enhancement of the
emission rate of the black hole as the number of extra dimensions projected
onto the brane increases, we compute the total flux and power emissivities,
for various values of $n$, by integrating Eqs.
(\ref{flux}) and (\ref{power}) over all energies. The results obtained
are displayed in Table 1. The relevant emissivities for different values
of $n$ have been normalized in terms of those for $n=0$. From the entries
of the table, we may easily see that both the flux of particles produced
and the amount of energy radiated per unit time by the black hole on the
brane are substantially enhanced, by orders of magnitude, as the number of
extra dimensions increases. 

%%%%%%%%%%%%%%%%%%%
\setcounter{table}{0}
\begin{center}
\TABLE{\begin{tabular}{c}
{\begin{tabular}{|c||c|c|c|c|c|c|c|c|}
\hline \hline {\rule[-3mm]{0mm}{8mm} }
 & $n=0$  & $n=1$ & $n=2$ &  $n=3$ 
 & $n=4$  & $n=5$ & $n=6$  & $n=7$ 
 \\[6pt]
\hline {\rule[-3mm]{0mm}{9mm} 
{\rm Flux}} & 1.0 & 4.75 & 13.0 & 27.4 & 49.3 & 79.9 & 121 & 172 \\
\hline {\rule[-3mm]{0mm}{9mm} 
{\rm Power}} & 1.0 & 8.94 & 36.0 & 99.8 & 222 & 429 & 749 & 1220 \\
\hline \hline 
\end{tabular}} \\[15mm]
{\small {\bf Table 1\,:} Flux and Power Emissivities for Scalar Fields
on the brane} \end{tabular}}
\end{center}
%%%%%%%%%%%%%%%%%%

\subsection{Spin 1/2 fields}

Unlike the case of scalar fields, the study of the emission of fields with
non-vanishing spin involves, in principle, the study of fields with more than
one component. Equation (\ref{radial}) depends on the helicity number $s$ 
which, upon taking different values, leads to the radial equation for different
components of the field. As mentioned in section 3.2, the upper and lower
components carry mainly the in-going and out-going parts, respectively, of
the field. Although knowledge of both components is necessary in order to
construct the complete solution for the emitted field, the determination
of either is more than adequate to compute the absorption coefficient
$\hat {\cal A}_j$, where $j$ is the total angular momentum number. For
example, if the in-going wave is known in the case of the emission of
fields with spin $s=1/2$, Eq. (\ref{absorption})
may be directly writen as \cite{kmr2}\cite{CL}
%%%%%%%%%%%
\begin{equation}
|\hat {\cal A}^{(1/2)}_j|^2=\Biggl|\frac{A_{in}^{(h)}}
{A_{in}^{(\infty)}}\Biggr|^2\,.
\end{equation}
%%%%%%%%%%
The above follows by defining the incoming flux of a fermionic field as the
radial component of the conserved current, 
$J^\mu=\sqrt{2}\,\sigma^\mu_{AB}\,\Psi^A\,\bar \Psi^B$, integrated over a
two-dimensional sphere and evaluated both at the horizon and infinity.

The greybody factor is again related to the aforementioned absorption
probability through Eq. (\ref{brane-loc}) with $\ell$ being replaced by $j$.
By numerically solving the radial equation
Eq. (\ref{radial}) and computing $\hat {\cal A}^{(1/2)}_j$, we obtain the
behaviour of the greybody factor, in terms of the energy parameter
$\omega r_H$ and number of extra dimensions $n$, depicted in Figure \ref{grey05}.
At low energies the greybody factor assumes, as in the case of scalar fields, 
a non-zero asymptotic value; this depends on the dimensionality of
spacetime and increases with increasing $n$. The enhancement of 
$\sigma^{(1/2)}_{\rm abs}(\omega)$ with $n$
in the low-energy regime persists up to intermediate values of $\omega r_H$,
after which the situation is reversed: as $n$ takes on larger values, the 
greybody factor becomes more and more suppressed. The complete analytic results
derived in \cite{kmr2} successfully describe both the low-energy behaviour and 
the existence of the `turning point'; however, as expected, they fail to
give accurate information for the high-energy regime. Figure \ref{grey05}
shows that at high energies the greybody factors for fermion fields oscillate
around the same asymptotic values [determined by the effective radius 
(\ref{effective})] as for scalar fields.

\FIGURE[b]{
\scalebox{0.5}{\rotatebox{-90}{\includegraphics[width=\textwidth]{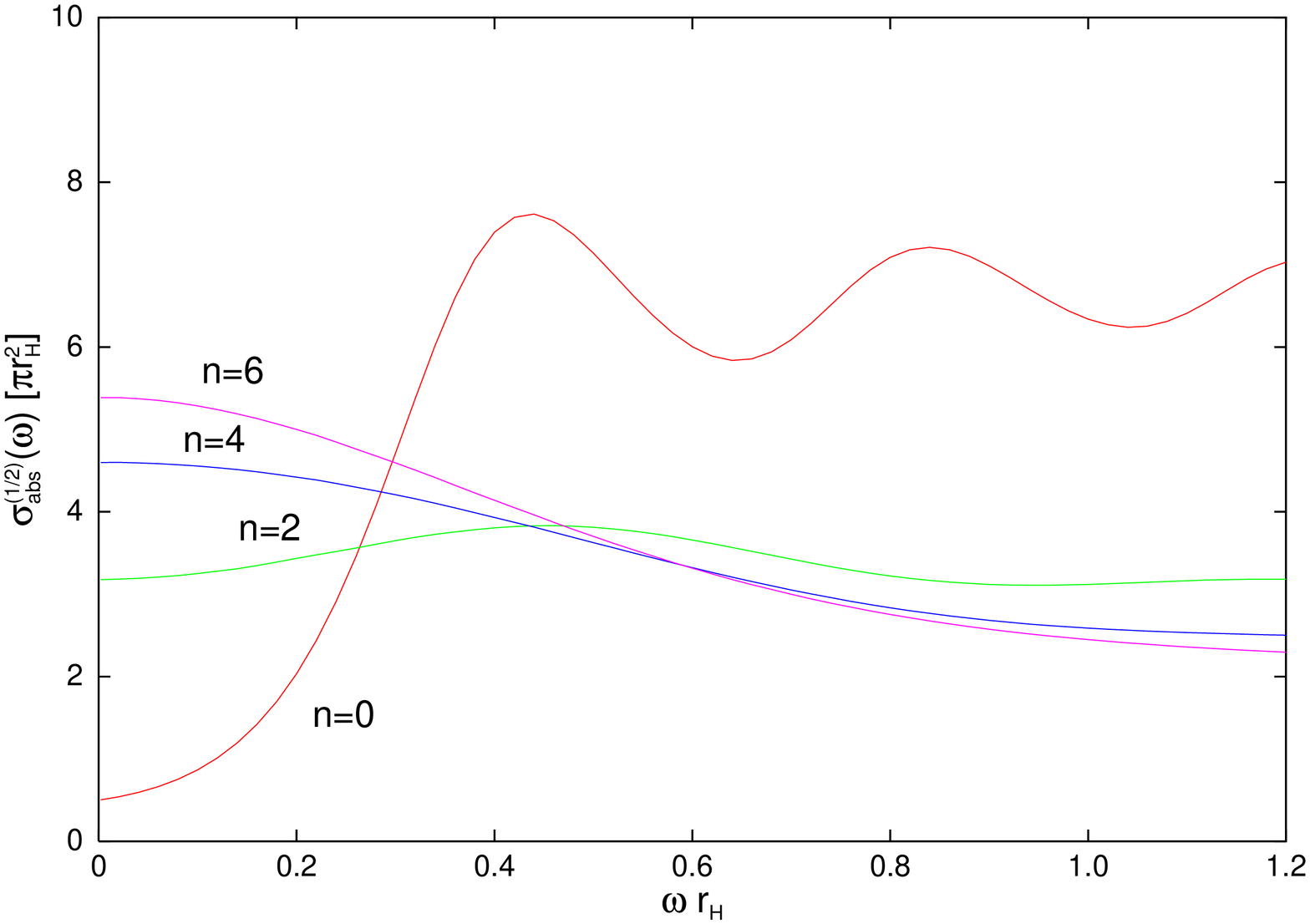}}}
\caption{Greybody factors for fermion emission on the brane from a $(4+n)$D
black hole.}
\label{grey05}
}

\FIGURE[t]{ \hspace{-0.8cm}
\scalebox{0.5}{\rotatebox{0}{\includegraphics[width=23cm, height=15cm]
{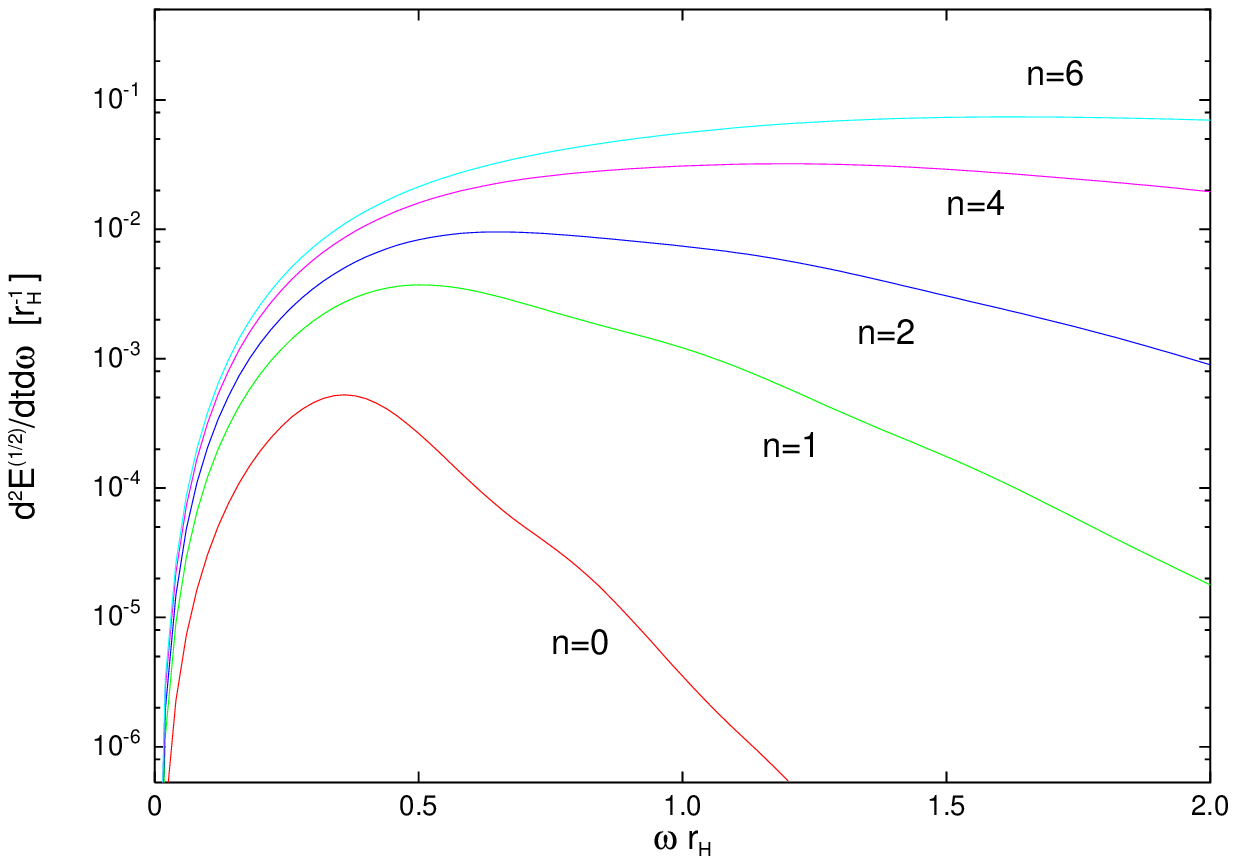}}}
\caption{Energy emission rates for fermions on the brane from a $(4+n)$D black hole.}
\label{fer-dec}
}
The energy emission rate for fermion fields on the brane, for various values
of $n$, is shown in Figure \ref{fer-dec}.
As $n$ increases, it is found to be significantly enhanced, both at low and 
high energies, mainly due to the increase in the temperature of the black
hole. The emission curves exhibit the same features as for the
emission of scalar fields, i.e. increase of the height of the peak by 
orders of magnitude and shift of the peak towards higher energies. Some
quantitative results regarding the enhancement of both the flux and power
spectra for the emission of fermions, as $n$ increases, are given in
Table 2. Once again, the enhancement in both spectra with $n$ is indeed
substantial, and even more important compared to the one for scalar
emission.

%%%%%%%%%%%%%%%%%%%
\begin{center}
\TABLE{\begin{tabular}{c}
\begin{tabular}{|c||c|c|c|c|c|c|c|c|}
\hline \hline {\rule[-3mm]{0mm}{8mm} }
 & $n=0$  & $n=1$ & $n=2$ &  $n=3$ 
 & $n=4$  & $n=5$ & $n=6$  & $n=7$ 
 \\[6pt]
\hline {\rule[-3mm]{0mm}{9mm} 
{\rm Flux}} & 1.0 & 9.05 & 27.6 & 58.2 & 103 & 163 & 240 & 335 \\
\hline {\rule[-3mm]{0mm}{9mm} 
{\rm Power}} & 1.0 & 14.2 & 59.5 & 162 & 352 & 664 & 1140 & 1830\\
\hline \hline 
\end{tabular}\\[15mm]
{\small {\bf Table 2\,:} Flux and Power Emissivities for Fermions 
on the brane} \end{tabular}}
\end{center}
%%%%%%%%%%%%%%%%%%

\subsection{Spin 1 fields}

\noindent
In the case of the emission of gauge boson fields, the incoming flux can be
computed by the $(tr)$-component of the energy-momentum tensor,
$T^{\mu\nu}=2 \sigma^{\mu}_{AA'} \sigma^{\nu}_{BB'} \Psi^{AB}\,\bar\Psi^{A'B'}$, integrated again over a two-dimensional sphere and evaluated at the horizon 
and infinity. By making use of the solution for the in-going wave, the 
following expression for the absorption probability, Eq. (\ref{absorption}),
is obtained \cite{kmr2}\cite{CL}:
%%%%%%%%%%%
\begin{equation}
|\hat {\cal A}^{(1)}_j|^2=\frac{1}{r_H^2}\,
\Biggl|\frac{A_{in}^{(h)}}{A_{in}^{(\infty)}}\Biggr|^2\,.
\end{equation}
%%%%%%%%%%

\FIGURE{
\scalebox{0.43}{\rotatebox{-90}{\includegraphics{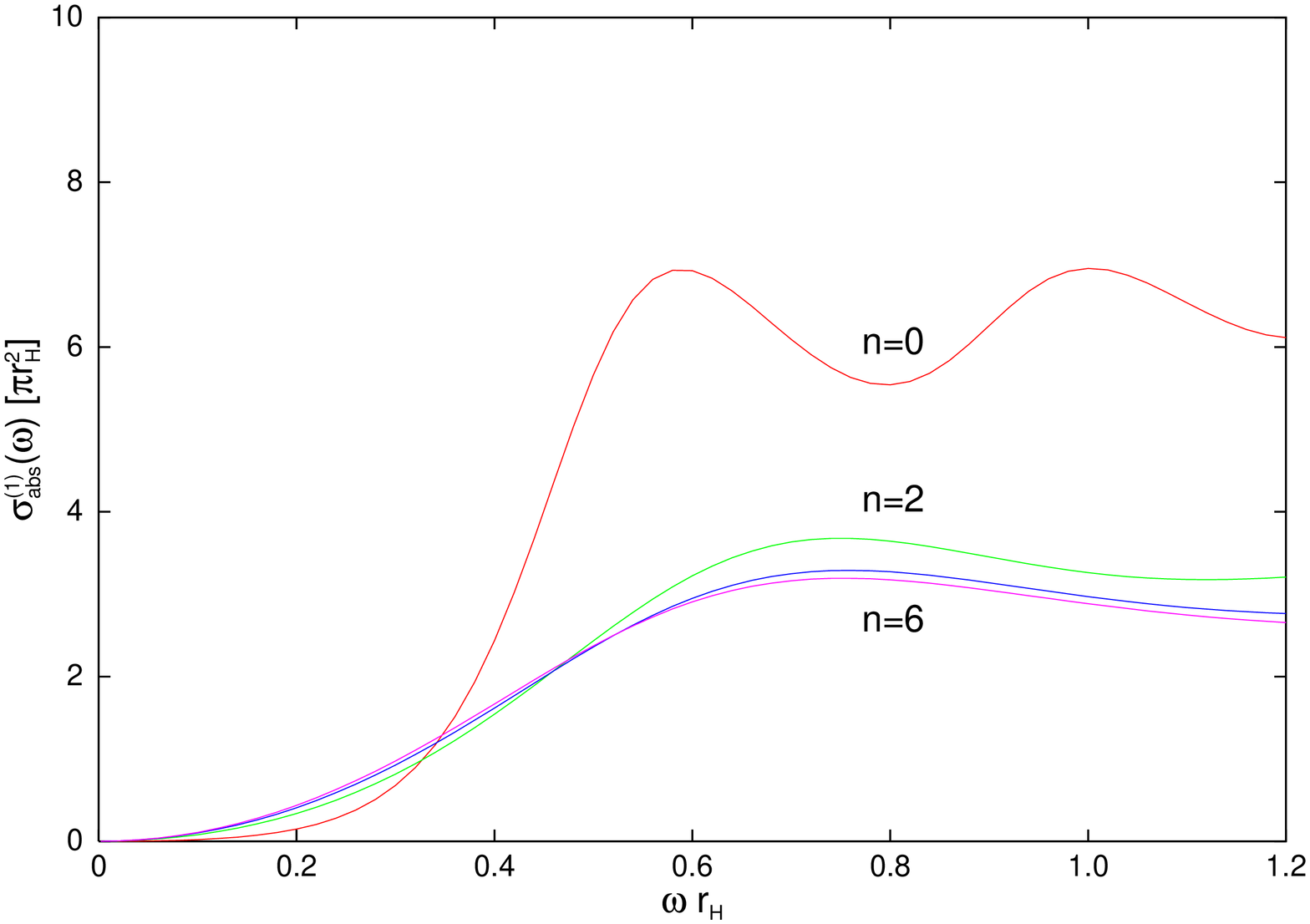}}}
\caption{Greybody factors for gauge boson emission on the brane from a
$(4+n)$D black hole.}
\label{grey1}
} 
\FIGURE{ \hspace*{-0.5cm}
\scalebox{0.5}{\rotatebox{0}{\includegraphics[width=23cm, height=15cm]
{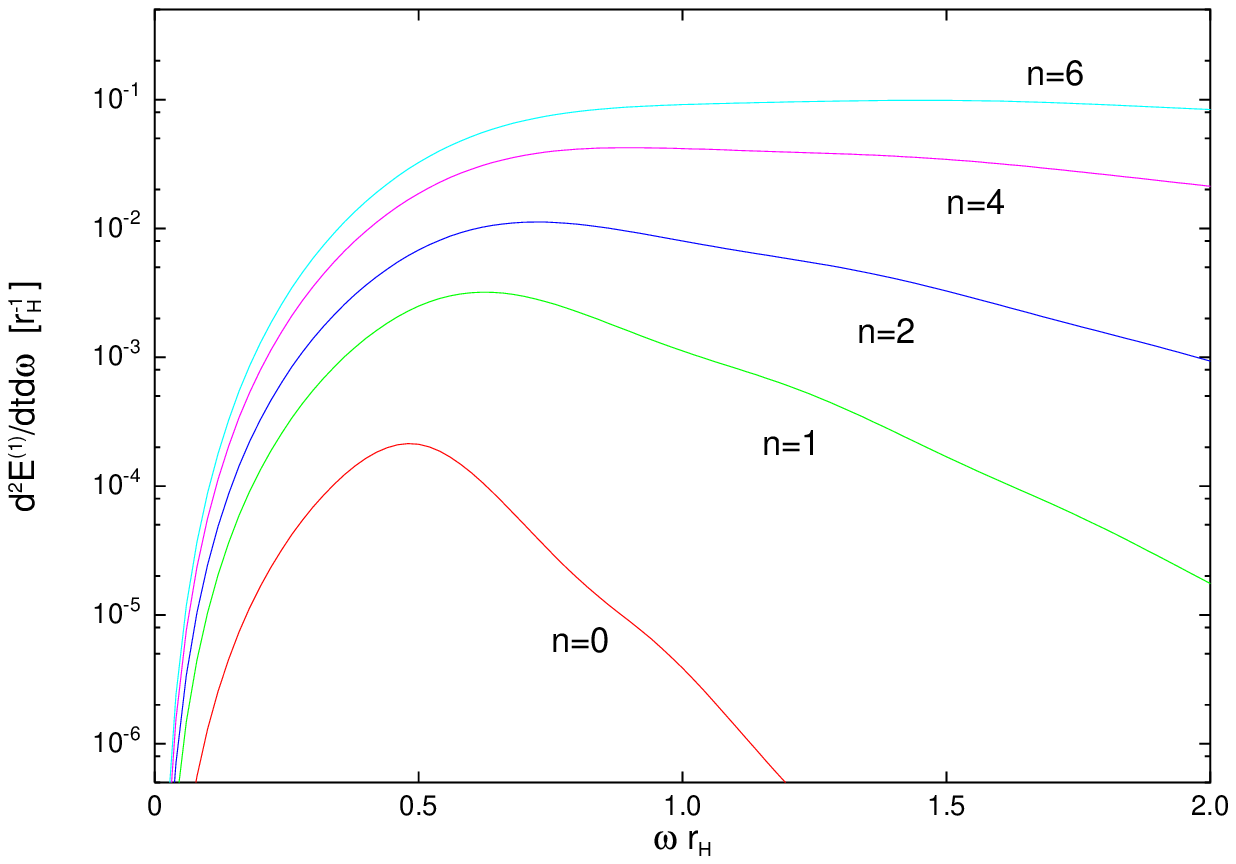}}}
\caption{Energy emission rates for gauge fields on the brane from a $(4+n)$D
black hole.}
\label{gauge-dec}
}

The exact results for the greybody factors and energy emission rates for 
gauge boson fields are given in Figures \ref{grey1} and \ref{gauge-dec}
respectively. A distinct feature of the greybody factor for gauge fields,
already known from the 4-dimensional case, is that it vanishes  when 
$\omega r_H \rightarrow 0$. The same behaviour is observed for every value of 
the 
number of extra dimensions. This result leads to the suppression
of the energy emission rate, in the low-energy regime, compared to the ones
for scalar and fermion fields.  Up to intermediate energies the greybody factors exhibit the same enhancement with increasing $n$ as in the case of
fermion fields, and as it was analytically shown in \cite{kmr2}. A similar
asymptotic behaviour, as in the previous cases, is observed in the high-energy
regime with each greybody factor assuming, after oscillation, the geometrical
optics value which decreases with increasing $n$. This result establishes the 
existence of a universal behaviour of all types of particles emitted by the 
black hole at high energies. This behaviour is independent of the particle 
spin but strongly dependent on the number of extra dimensions projected onto 
the brane. 

We may finally obtain, as in the previous cases, the total flux and power
emissivities for the emission of gauge fields on the brane, in terms of the
number of extra dimensions $n$. The exact results obtained by numerically
integrating over all energies are given in Table 3.
As anticipated, the same pattern of enhancement with $n$ is also observed 
for the emission of gauge bosons. It is worth noting that 
the enhancement observed in this case is the largest amongst all types of
particles --- this result points to the dominance of the emission of gauge
bosons over other types of particles in models with large values of $n$,
as we will shortly see.

%%%%%%%%%%%%%%%%%%%
\begin{center}
\TABLE{\begin{tabular}{c}
\begin{tabular}{|c||c|c|c|c|c|c|c|c|}
\hline \hline {\rule[-3mm]{0mm}{8mm} }
 & $n=0$  & $n=1$ & $n=2$ &  $n=3$ 
 & $n=4$  & $n=5$ & $n=6$  & $n=7$ 
 \\[6pt]
\hline {\rule[-3mm]{0mm}{9mm} 
{\rm Flux}} & 1.0 & 19.2 & 80.6 & 204 & 403 & 689 & 1070 & 1560 \\
[6pt]
\hline {\rule[-3mm]{0mm}{9mm} 
{\rm Power}} & 1.0 & 27.1 & 144  & 441 & 1020 & 2000 & 3530 & 5740 \\
\hline \hline 
\end{tabular}\\[15mm]
{\small {\bf Table 3\,:} Flux and Power Emissivities for Gauge Fields
on the brane} \end{tabular}}
\end{center}
%%%%%%%%%%%%%%%%%%

\subsection{Relative Emissivities for different species}

It would be interesting to investigate how the relative numbers of scalars,
fermions and gauge bosons, emitted by the black hole on the brane, change
as the number of extra dimensions projected onto the brane varies. In other
words, we would like to know what type of particles the black hole prefers
to emit, for different values of $n$, and what part of the total energy
each particular type of particle carries away during emission. 

\FIGURE[t]{
\scalebox{0.5}{\rotatebox{0} 
{\includegraphics[width=14cm, height=13cm]{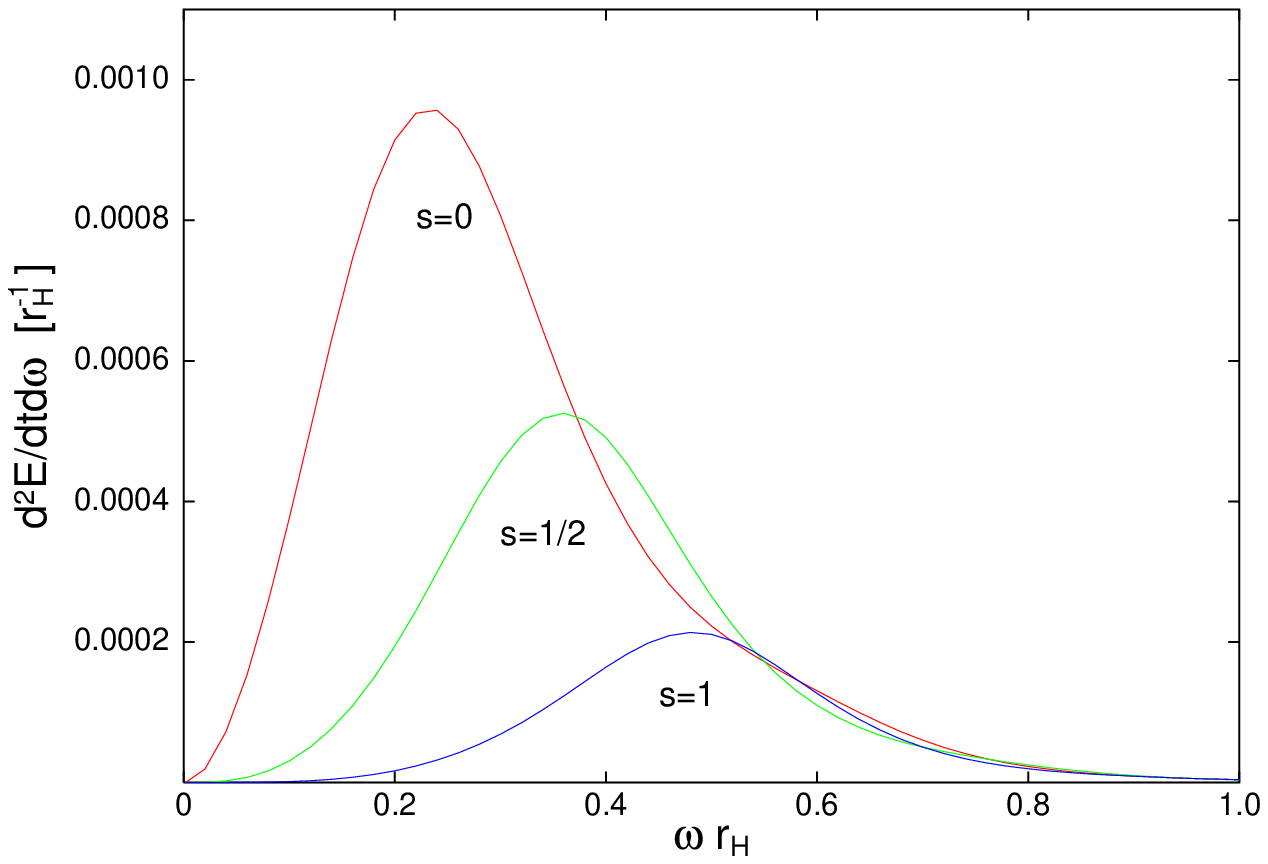}}}
\caption{Greybody factors for gauge boson emission from non-rotating black 
holes.}
\scalebox{0.5}
{\rotatebox{0}{\includegraphics[width=14cm, height=13cm]{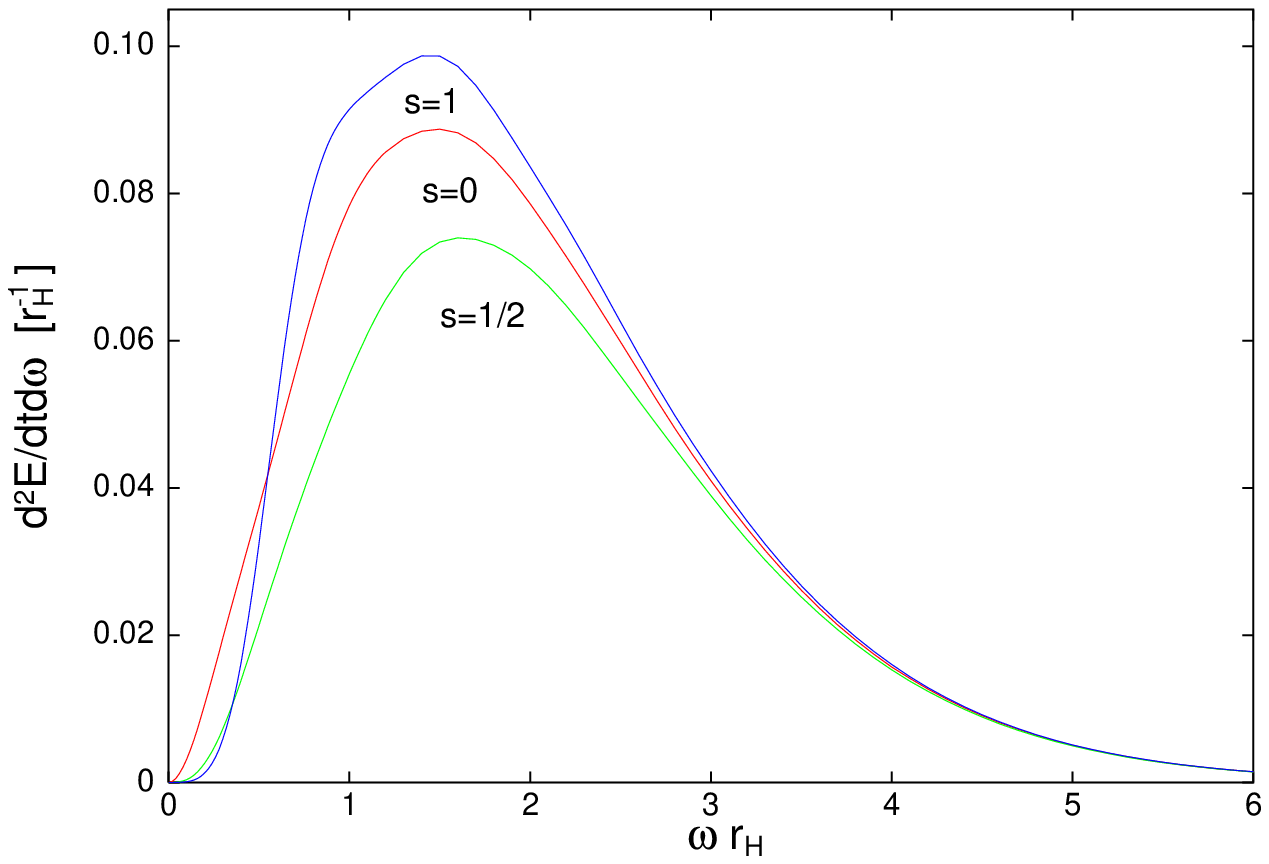}}}
\caption{{\bf (a)\,:} Energy emission rates for the emission of scalars, fermions and 
gauge bosons on the brane for $n=0$. {\bf (b)\,:} The same, but for $n=6$.}
\label{relan6}
} 

Comparing the energy emission rates (computed in the previous sub-sections),
for different types of particles and for fixed $n$, can give us the
qualitative behaviour. Figures \ref{relan6}(a) and \ref{relan6}(b) show, with 
a linear scale, the above quantities for $n=0$ and $n=6$ respectively --- note 
that these two figures summarize very clearly the
effects discussed in the previous sub-sections, i.e. the orders-of-magnitude
enhancement of the emission rates and the Wien's displacement of the peak
to the right, as $n$ increases. Figure \ref{relan6}(a) reveals
that, in the absence of any extra dimensions, most of the energy of
the black hole emitted on the brane is in the form of scalar particles;
the next most important are the fermion fields, and less significant are the 
gauge bosons. As
$n$ increases, the emission rates for all species are enhanced but not
at the same rate. Figure \ref{relan6}(b) clearly shows that, for a 
large number of extra dimensions, the most effective `channel' during
the emission of brane localized modes is that of gauge bosons;
the scalar and fermion fields follow second and third respectively.
The change in the flux spectra, i.e. in the number of particles produced
by the black hole on the brane, for each species is similar as $n$ increases. 

\setcounter{table}{3}
\DOUBLETABLE[b]{\begin{tabular}{|c||c|c|c|}
\hline \hline {\rule[-3mm]{0mm}{8mm} }
& $s=0$ & $s=\frac{1}{2}$ & $s=1$ \\[6pt]
\hline
$n=0$&1.0&0.37&0.11\\
$n=1$&1.0&0.70&0.45\\
$n=2$&1.0&0.77&0.69\\
$n=3$&1.0&0.78&0.83\\
$n=4$&1.0&0.76&0.91\\
$n=5$&1.0&0.74&0.96\\
$n=6$&1.0&0.73&0.99\\
$n=7$&1.0&0.71&1.01\\
\hline 
Blackbody&1.0&0.75&1.0\\
\hline \hline
\end{tabular}}
{\begin{tabular}{|c||c|c|c|}
\hline \hline {\rule[-3mm]{0mm}{8mm} }
& $s=0$ & $s=\frac{1}{2}$ & $s=1$ \\[6pt]
\hline
$n=0$&1.0&0.55&0.23\\
$n=1$&1.0&0.87&0.69\\
$n=2$&1.0&0.91&0.91\\
$n=3$&1.0&0.89&1.00\\
$n=4$&1.0&0.87&1.04\\
$n=5$&1.0&0.85&1.06\\
$n=6$&1.0&0.84&1.06\\
$n=7$&1.0&0.82&1.07\\
\hline 
Blackbody&1.0&0.87&1.0\\
\hline \hline
\end{tabular}}
{Flux emission ratios \label{fratios}}
{Power emission ratios \label{pratios}}

In order to quantify the above behaviour, we computed the relative emissivities
for scalars, fermions and gauge bosons emitted on the brane by integrating
the flux (\ref{flux}) and energy (\ref{power}) emission spectra, for different
types of particles, over all energies. The relative emissivities obtained in
this way are shown in Tables \ref{fratios} and \ref{pratios} (they are 
normalized to the
scalar values). Note that the ratios for $n\geq1$ are available for
the first time in the literature as a result of this numerical work, while
the $n=0$ results would appear to be the most accurate ones available. 
The entries in these tables reflect the qualitative behaviour
discussed above for some extreme values of the number of extra dimensions.
For $n=0$, the scalar fields are indeed the type of particles which are
most commonly produced and the ones which carry away most of the energy of
the black hole emitted on the brane; the fermion and gauge fields carry
approximately 1/2 and 1/4, respectively, of the energy emitted in the form 
of the scalar fields, and their fluxes are only 1/3 and 1/10 of the scalar 
flux. For intermediate values of $n$, the fermion and
gauge boson emissivities have been considerably enhanced compared to
the scalar one and have become of approximately the same magnitude --- e.g.
for $n=2$ the amount of energy spent by the black hole for the emission
of fermions and gauge bosons is exactly the same, although the net number
of gauge bosons is still sub-dominant. For large values of $n$, the situation
is reversed: the gauge bosons dominate both flux and power spectra, with
the emission of fermions being the least effective `channel' both in
terms of number of particles produced and energy emitted. We remind the
reader that the above results refer to the emission of individual scalar,
fermionic or bosonic degrees of freedom and not to elementary particles.

We may thus conclude that not only the magnitude but also the type of flux
and power spectra produced by a small, higher-dimensional black hole strongly
depends on the number of extra dimensions projected onto the brane. Therefore,
upon detecting the Hawking radiation from such objects, the above distinctive
feature could serve as an alternative way to determine the number of extra
dimensions that exist in nature.

%%%%%%%%%%%%%%%%%%%%%%%%%%%%%%%%%%%%%%%%%%%%%%%%%%%%%%%%%%%%%%%%%%%%%%%%

\section{Emission in the Bulk}

An extremely important question regarding the emission of particles by
a higher-dimensional black hole is how much of this energy is radiated
onto the brane and how much is lost in the bulk. In the former case,
the emitted particles are zero-mode gravitons and scalar fields as 
well as Standard Model fermions and gauge bosons, while in the latter
case all emitted energy is in the form of massive Kaluza-Klein gravitons
and, possibly, scalar fields. In \cite{emparan}, it was shown that the whole
tower of KK excitations of a given particle carries approximately the
same amount of energy as the massless zero-mode particle emitted on the
brane. Combining this result with the fact that many more types of particles
live on the brane than in the bulk, it was concluded that most of the
energy of the black hole goes into brane modes. The results obtained
in \cite{emparan} were only approximate since the dependence of the greybody
factor on the energy of the emitted particle was ignored and the
(low-energy valid) geometric expression for the area of the horizon was
used instead.  

In order to provide an accurate answer to the question of how much energy
is emitted into the bulk compared to on the brane, it is imperative
that the dependence of the greybody factor on both the energy and number of 
extra dimensions is taken into account (for a similar but incomplete ---
since the dependence on the energy was again ignored --- argument in this
direction, see \cite{cavaglia}). In this section, we first thoroughly
investigate the
emission of scalar modes in the bulk and produce exact numerical results
for the behaviour of greybody factors and energy emission rates in terms
of the energy and number of extra dimensions. Subsequently, we address
the above question and provide a definite answer by computing, for different 
values of $n$, the total bulk-to-brane relative emissivities. 

\smallskip
\subsection{Greybody Factors and Emission Rates}

In this section, we turn to the investigation of the emission of bulk modes
from a higher-dimensional black hole. This analysis is relevant for gravitons
and scalar fields, and requires knowledge of the solutions of the
corresponding equations of motion in the bulk. Here, we will focus on the
case of scalar fields for which the bulk equation is known --- the emission
of bulk scalar modes was previously studied analytically, in the low-energy
regime, in \cite{kmr1}.

A scalar field propagating in the background of a higher-dimensional,
non-rotating, Schwarzschild-like black hole, whose line-element is given by
Eq. (\ref{metric-D}), satisfies the following equation of motion \cite{kmr1}
%%%%%%%%%%%%
\begin{equation}
\frac{h(r)}{r^{n+2}}\,\frac{d \,}{dr}\,
\biggl[\,h(r)\,r^{n+2}\,\frac{d R}{dr}\,\biggr] +
\biggl[\,\om^2 - \frac{h(r)}{r^2}\,\ell\,(\ell+n+1)\,\biggr] R =0 \, .
\label{scalareqn}
\end{equation}
%%%%%%%%%%%

As in the case of the emission of particles on the brane, the determination
of the greybody factor for emission in the bulk demands solving the above
equation over the whole radial domain. The exact solution for the radial
function must interpolate between the near-horizon and far-field asymptotic
solutions, given by
%%%%%%%%%
\begin{equation}
\label{near-bulk}
R^{(h)}=A_{in}^{(h)}\,e^{-i\omega r^{*}}+ A_{out }^{(h)}\,e^{i\omega r^{*}},
\end{equation}
%%%%%%%%%%
and
%%%%%%%%%%%
\begin{equation}
\label{far-bulk}
R^{(\infty)}=A_{in}^{(\infty)}\,\frac{e^{-i\omega r}}{\sqrt{r^{n+2}}}+
A_{out}^{(\infty)}\,\frac{e^{i\omega r}}{\sqrt{r^{n+2}}}\,,
\end{equation}
%%%%%%%%%%%%%
respectively. We impose again the boundary condition that no out-going
solution should exist near the horizon of the black hole, and therefore we
set $A_{out}^{(h)}=0$. The solution at infinity comprises, as usual, both
in-going and out-going modes. 

The expression for the absorption probability $|\tilde {\cal A}_\ell|^2$ may
then be derived either by using Eq. (\ref{scalars}) or by calculating directly
the ratio $|A_{in}^{(h)}/A_{in}^{(\infty)}|$ --- note
that, henceforth, quantities with a tilde denote bulk quantities (as opposed
to brane quantities which carry a hat). The corresponding greybody factor
$\tilde \si_\ell(\om)$ may then be determined by using the relation
(\ref{greybody}). The dimensionality of the greybody factor changes as the
number of extra dimensions $n$ varies; therefore, in order to be able to
compare its values for different $n$, we normalize its expression to
the area of the horizon of the $(4+n)$-dimensional black hole. Thus, we
rewrite Eq. (\ref{greybody}) in the form
%%%%%%%%%%%%
\begin{equation}
\tilde \si_\ell(\om) = \frac{2^{n}}{\pi}\,
\Ga\Bigl[\frac{n+3}{2}\Bigr]^2\,
\frac{\tilde A_H}{(\om r_H)^{n+2}}\, \tilde N_\ell\,
|\tilde {\cal A}_\ell|^2\,,
\label{greyb}
\end{equation}
%%%%%%%%%%%%
where $\tilde N_\ell$ is the multiplicity of states corresponding to the
same partial wave $\ell$, given for a $(4+n)$-dimensional spacetime by
%%%%%%%%%
\begin{equation}
\tilde N_\ell= \frac{(2\ell+n+1)\,(\ell+n)!}{\ell! \,(n+1)!}\,,
\label{bulk-mult}
\end{equation}
%%%%%%%%%
and $\tilde A_H$ is the horizon area in the bulk defined as
%%%%%%%%%
\begin{eqnarray}
\tilde A_H &=& 
r_H^{n+2}\,\int_0^{2 \pi} \,d \varphi \,\prod_{k=1}^{n+1}\,
\int_0^\pi\,\sin^k\theta_{n+1}\,
\,d\sin\theta_{n+1} \nonumber \\[2mm]
&=& r_H^{n+2}\,(2\pi)\,\prod_{k=1}^{n+1}\,\sqrt{\pi}\,\,\frac{\Ga[(k+1)/2]}
{\Ga[(k+2)/2]}\nonumber \\[2mm]
&=& 
r_H^{n+2}\,(2\pi)\,\pi^{(n+1)/2}\,\Ga\Bigl[\frac{n+3}{2}\Bigr]^{-1}\,.
\end{eqnarray}
%%%%%%%%%%

Equation (\ref{greyb}) allows us to compute the low-energy limit of the
greybody factor once the corresponding expression of the absorption
coefficient is determined. Analytic results for $\tilde {\cal A}_\ell$
were derived in \cite{kmr1} by solving Eq. (\ref{scalareqn}) in the two
asymptotic regimes, the near-horizon and far-field, and matching them
in an intermediate zone. It was found that the low-energy
expression of the absorption coefficient, for $\ell=0$, has the form
%%%%%%%%%%%%%
\begin{equation}
|\tilde {\cal A}_0|^2 = \biggl(\frac{\omega r_H}{2}\biggl)^{n+2}
\,\frac{4 \pi}{\Gamma[(n+3)/2]^2} + ... \,,
\end{equation}
%%%%%%%%%%%%%%%
where the dots denote higher-order terms in the power-expansion
in $\omega r_H$. These terms, as well as the corresponding
expressions of $\tilde {\cal A}_\ell$ for higher partial waves, vanish
quickly in the limit $\omega r_H \rightarrow 0$, 
leaving the above term as the dominant one. Substituting
into Eq. (\ref{greyb}), we easily see that, in the low-energy regime,
the greybody factor is given by the area $\tilde A_H$ of the horizon
of the black hole. This behaviour is similar to the 4-dimensional case;
however, note that, in this case, the area of the horizon changes as $n$
varies. 

In the high-energy regime, we anticipate recovering the equivalent of
the $(4+n)$-dimensional geometrical optics limit. In four dimensions, the
low-energy limit for the greybody factor, $4\pi r_H^2$, goes over to the
geometrical optics value $\pi r_c^2$ at high energy.  This has led to
the na\"{\i}ve generalization that, in an arbitrary number of dimensions, the
high-energy expression for the greybody factor will be approximately
$\Omega_{n+2}\,r_c^{n+2}/4$, where $\Omega_{n+2}$ is the volume element
of the $(n+2)$-dimensional sphere. We will shortly see that this is in
fact an over-estimate of the value of the greybody factor in the
high-energy regime. As in four dimensions, we will assume that, for large
values of the energy of the scattered particle, the greybody factor
becomes equal to the area of an absorptive body of radius $r_c$ which is
projected on a plane parallel to the one of the orbit of the moving particle
\cite{MTW}. According to Ref. \cite{emparan}, the value of the effective
radius $r_c$ remains the same both for bulk and brane particles and it
is given by Eq. (\ref{effective}). The area of the absorptive body depends
strongly on the dimensionality of spacetime and its calculation demands
setting one of the azimuthal angles equal to $\pi/2$. A careful calculation
reveals that the `projected' area is given by
%%%%%%%%%%%%
\begin{equation}
\tilde A_p = \frac{2 \pi}{(n+2)}\,\frac{\pi^{n/2}}{\Gamma[(n+2)/2]}\,
r_c^{n+2} = \frac{1}{n+2}\,\Omega_{n+1}\,r_c^{n+2}\,.
\end{equation}
%%%%%%%%%%%%
The above relation reduces to the usual 4-dimensional result
$\tilde A_p = \pi r_c^2$, for $n=0$, while it leads to values reduced by
50\%, compared to the \,na\"{\i}ve \,expression $\Omega_{n+2}\,r_c^{n+2}/4$,
for higher values of $n$. Assuming that the greybody factor at high
energies becomes equal to the absorptive area $\tilde A_p$ of radius $r_c$,
we may explicitly write:
%%%%%%%%%%%%%%
\begin{eqnarray}
\tilde \si_\ell(\om) &=& \frac{1}{n+2}\,\frac{\Omega_{n+1}}{\Omega_{n+2}}\,
\biggl(\frac{r_c}{r_H}\biggr)^{n+2}\,\tilde A_H \nonumber \\[3mm] &=&
\frac{1}{\sqrt{\pi}\,(n+2)}\,\frac{\Gamma[(n+3)/2]}{\Gamma[(n+2)/2]}\,
\biggl(\frac{n+3}{2}\biggr)^{(n+2)/(n+1)}\,
\biggl(\frac{n+3}{n+1}\biggr)^{(n+2)/2}\,\tilde A_H\,.
\label{high}
\end{eqnarray}
%%%%%%%%%%%%
In the above, we have used the same normalization, in terms of the area of
the $(4+n)$-dimensional horizon, as in the low-energy regime.

Turning now to the numerical analysis, we may find the expressions of the
greybody factors, for various values of $n$ and for all energy regimes,
by using Eq. (\ref{greyb}) and the exact numerical results for the absorption
coefficients. Their behaviour is shown in Figure \ref{grey0-bulk}. As
it was anticipated after the above discussion, the normalized greybody 
factors, in the low-energy regime, tend to unity for all values of $n$,
as each one of them adopts the value of the area of the black hole horizon
to which it has been normalized. As in the case of the emission of 
scalar fields on the brane, the greybody factors are suppressed in the 
low-energy regime, as $n$ increases, while they start oscillating at 
intermediate energies and finally adopt their asymptotic high-energy
limit. A simple numerical analysis shows that the na\"{\i}ve expression
$\Omega_{n+2}\,r_c^{n+2}/4$ fails to describe the high-energy asymptotic
limits for all values of $n$ larger than zero. On the other hand, the
expression (\ref{high}) gives asymptotic values which are much closer to the 
ones depicted in Figure \ref{grey0-bulk}, but these values still deviate
from the exact ones as $n$ increases. A more sophisticated analysis is
thus necessary in order to determine the exact high-energy limit for emission
in the bulk which might lead either to the reconsideration of the
absorptive-area argument for $n \geq 1$, or to the introduction of a
correcting term that suppresses the asymptotic value (\ref{high})
as $n$ increases.

\FIGURE[t]{
\scalebox{0.5}{\rotatebox{0}{\includegraphics[width=22cm, height=15.4cm]
{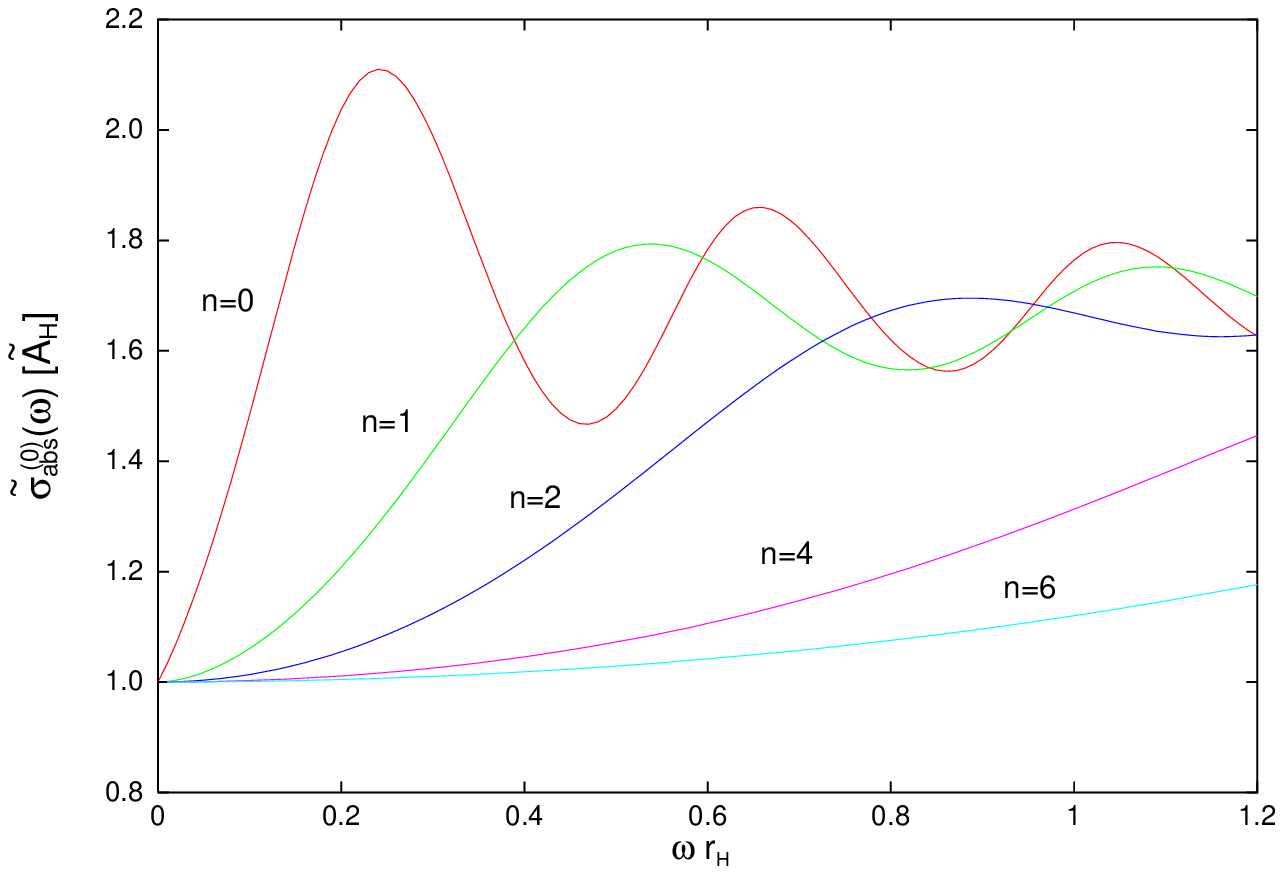}}}
\caption{Greybody factors for scalar emission in the bulk from a $(4+n)$D
black hole.}
\label{grey0-bulk}
}

\FIGURE{
\scalebox{0.5}{\rotatebox{0}{\includegraphics[width=23cm, height=15cm]
{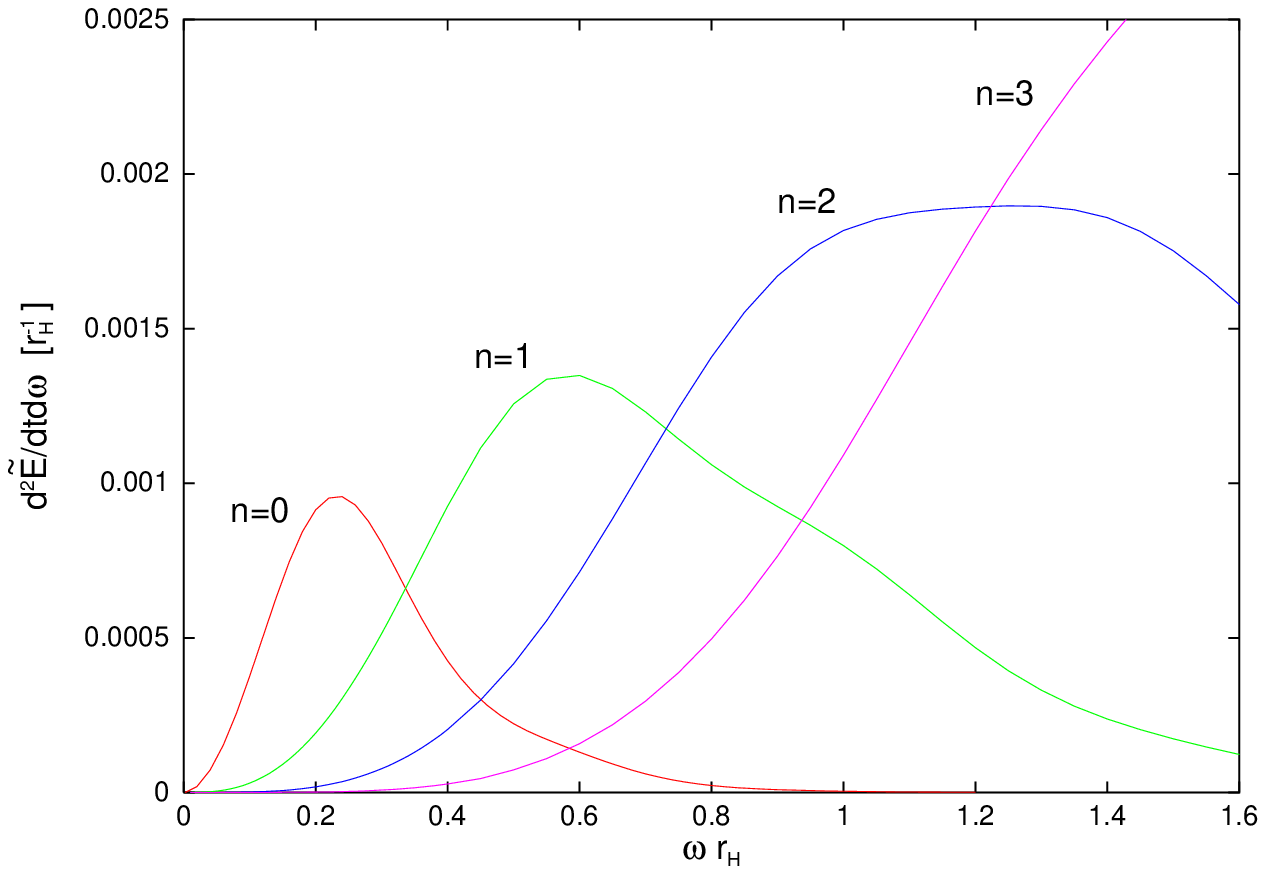}}}
\caption{Energy emission rates for scalar fields in the bulk from a $(4+n)$D
black hole.}
\label{rate-bulk}
}

In general, the suppression of the greybody factor for bulk emission at low
energies is milder than the one for brane emission. This, however, does not
lead to higher emission rates for the bulk modes compared to those for brane
modes: the integration over the phase-space in Eq. (\ref{power})
involves powers of $(\om r_H)$ which cause an increasingly
suppressive effect, in the low-energy regime, as $n$ increases. Nevertheless,
the increase in the temperature of the black hole, which is again given by
$T_{BH}=(n+1)/4 \pi r_H$, eventually overcomes the decrease in the value of
the greybody factor and causes the enhancement of the emission rate with $n$ at
high energies. The behaviour of the differential energy emission rates as a
function of the energy parameter $\om r_H$ and for some indicative numbers
of extra dimensions is depicted in Figure \ref{rate-bulk}. After the
aforementioned suppression at the low-energy regime as $n$ increases,
the energy emission rates soon become enhanced, with the peak of the curve
becoming higher and corresponding to larger values of $\om r_H$. It is
worth noting that the full analytic results, which may be derived from the
analysis of Ref. \cite{kmr1}, successfully describe the low-energy behaviour
of both the greybody factors and the energy emission rates for the emission
of scalar fields in the bulk.

\smallskip
\subsection{Bulk-to-Brane Relative Emissivities}

In this section, we perform an analysis aiming at providing an answer
to the question of the relative bulk-to-brane emissivity.  We evaluate
the differential energy emission rates in the bulk and on the brane, and
compare the two quantities for different numbers of extra dimensions. 

Equation (\ref{power}) for energy emission in the bulk may be alternatively
written, in terms of the absorption coefficient, as
%%%%%%%%%%%%%
\begin{equation}
\frac{d \tilde E(\om)}{dt} = 
\sum_{\ell} \tilde N_\ell\, |\tilde {\cal A}_\ell|^2\,{\om  \over
\exp\left(\om/T_{BH}\right) - 1}\,\,\frac{d \om}{2\pi}\,.
\label{alter-bulk}
\end{equation}
%%%%%%%%%%%%%
The above must be compared with the corresponding expression for the emission
of brane-localized modes given by
%%%%%%%%%%%%%
\begin{eqnarray}
\frac{d \hat E (\om)}{dt} =  \sum_{\ell} \hat N_\ell\, |\hat {\cal  A}_\ell|^2\,
{\om  \over \exp\left(\om/T_{BH}\right) - 1}\,\,\frac{d \om}{2\pi}\,,
\label{emission-br}
\end{eqnarray}
%%%%%%%%%%%%%
where $\hat N_\ell =2\ell +1$.
Since both bulk and brane modes `feel' the same temperature, the relative
bulk-to-brane ratio of the two energy emission rates will be simply given
by the expression 
\FIGURE{
\scalebox{0.5}{\rotatebox{0}{\includegraphics[width=23cm, height=15cm]
{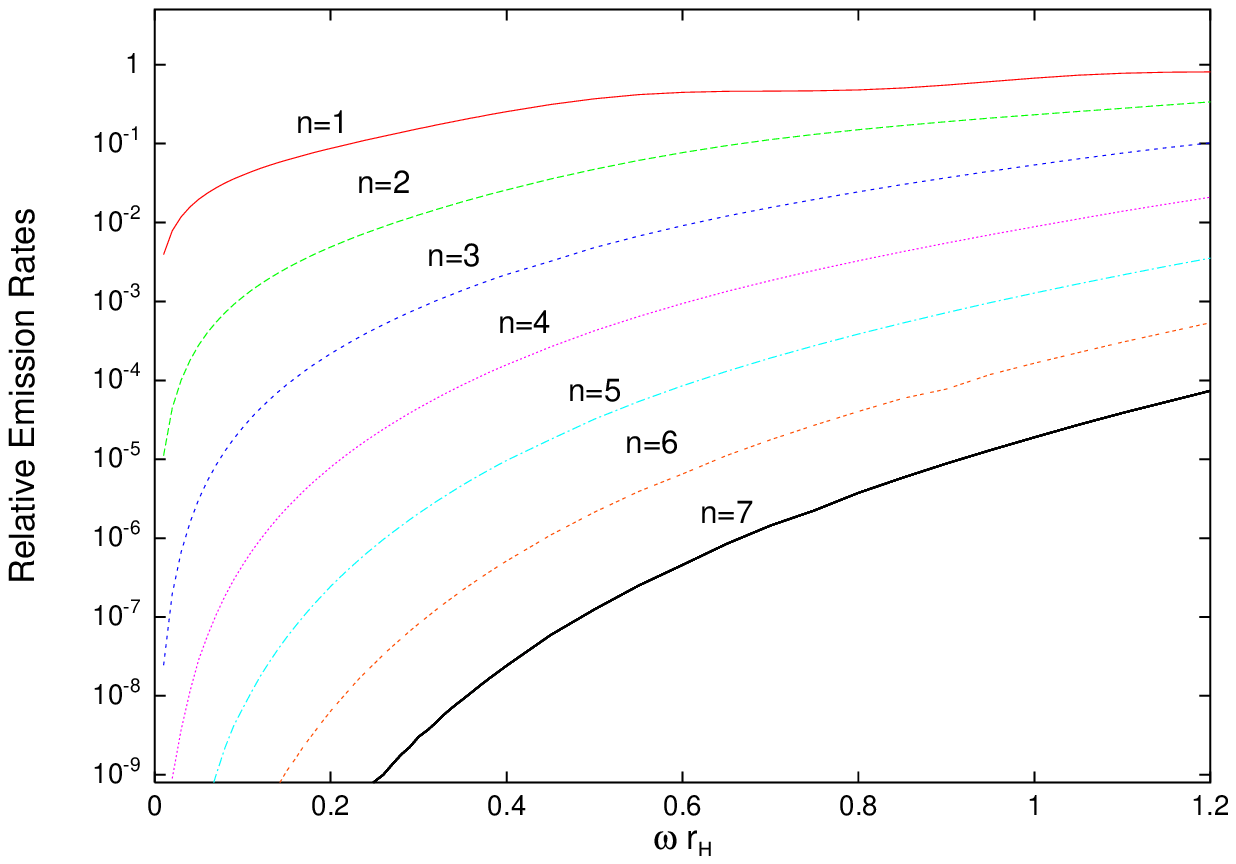}}}
\caption{Bulk-to-Brane energy emission rates for scalar fields from a
$(4+n)$D black hole.}
\label{bb-ratio}}
%%%%%%%%%%%%%
\begin{equation}
\frac{d \tilde E/dt}{d \hat E/dt} = \frac{\sum_{\ell} \tilde N_\ell\,
|\tilde {\cal A}_\ell|^2}{\sum_{\ell} \hat N_\ell\,
|\hat {\cal A}_\ell|^2}\,,
\label{ratio}
\end{equation}
%%%%%%%%%%%%%
and will depend on the scaling of the multiplicities of states and the
absorption coefficients\,\footnote{Note that the absorption coefficients
related to the greybody factors through Eq. (\ref{greybody}), with a
multiplicative coefficient that depends both on $\omega r_H$ and $n$,
might have a completely different behaviour from the greybody factors
themselves.} with $n$. As is clear from Eq. (\ref{bulk-mult}), the
multiplicity of bulk modes $\tilde N_\ell$ increases quickly for increasing
$n$, while $\hat N_\ell$ remains the same. However, it turns out that the
enhancement of the absorption probability $|\hat {\cal A}_\ell|^2$ for
brane emission, as $n$ increases, is in total considerably greater than
the one for bulk emission. This leads to the dominance of the emission of
brane-localized modes over bulk modes, which as we will see becomes stronger
for intermediate values of $n$. The behaviour of this ratio is depicted in
Figure \ref{bb-ratio}. We observe that, in the low-energy regime, the ratio
is suppressed, for large values of $n$, by many orders of magnitude, compared
to the value of unity for $n=0$. In the high-energy regime on the other hand,
the suppression becomes smaller and the ratio seems to approach unity.
A more careful examination reveals that, in fact, the bulk modes dominate
over the brane modes in a limited high-energy regime that becomes broader
as $n$ increases.

A definite conclusion regarding the relative amount of energy which is
emitted in the two `channels' --- bulk and brane --- can only be drawn if
the corresponding total energy emissivities can be computed. By integrating
the areas under the bulk and brane energy emission rate curves, for fixed
value of $n$, and taking their ratio we were able to determine the relative
energy emission rates. The results obtained, for values of $n$ from 1 to 7,
are given in Table 6.

%%%%%%%%%%%%%%%%%%%
\begin{center}
\TABLE{\begin{tabular}{|c||c|c|c|c|c|c|c|c|}
\hline \hline {\rule[-3mm]{0mm}{8mm} }
 & $n=0$ & $n=1$  & $n=2$  & $n=3$ & $n=4$ & $n=5$ & $n=6$ & $n=7$ \\[6pt]
\hline {\rule[-3mm]{0mm}{9mm} 
{\rm Bulk/Brane}} & 1.0 & 0.40 & 0.24 & 0.22 & 0.24 & 0.33 & 0.52 & 0.93\\
\hline \hline 
\end{tabular}}
\vspace*{-6mm}
\caption{Relative Bulk-to-Brane Energy Emission Rates for Scalar Fields
\label{bb-ratios}}
\end{center}
%%%%%%%%%%%%%%%%%%

From the entries of the above Table, it becomes clear that the emission of
brane-localized scalar modes is indeed dominant, in terms of the energy
emitted, for all values of $n$ greater than zero and up to 7. As $n$ increases, the
ratio of bulk to brane emission gradually becomes smaller, and becomes
particularly suppressed for intermediate values of the number of extra
dimensions, i.e. $n=2,3,4$ and 5; in these cases, the total energy emitted
in the bulk varies between 1/3 and 1/4 approximately of that emitted on
the brane. As $n$ increases further, the high-energy dominance of the bulk
modes, mentioned above, gives a boost to the value of the bulk-to-brane
ratio thus causing its increase --- nevertheless, the energy ratio never
exceeds unity. This means that most of the energy of the higher-dimensional
black hole, in the `scalar' channel, is emitted directly on the brane, in
the form of zero-mode scalar fields instead of modes in the bulk. 

The above analysis provides exact, accurate results for the energy emission
rates for brane and bulk scalar modes and gives considerable support to
earlier, more heuristic, arguments \cite{emparan}, according to which a
$(4+n)$-dimensional black hole emits mainly brane modes. A complete
confirmation demands performing a similar analysis for the emission of
gravitons and the results will be reported elsewhere.

\section{Conclusions}

The revival of the idea of the existence of extra space-like dimensions
in nature has led to the formulation of theories that allow for a 
significantly lower energy scale at which gravity becomes strong. This
has opened the way for the proposal of the creation of miniature
higher-dimensional black holes during collisions of energetic particles
in the earth's atmosphere or at ground-based particle colliders.
These black holes have a temperature that depends on details of the
higher-dimensional background, such as the horizon radius and the
number of spatial dimensions, and, upon implementation of quantum effects,
emit Hawking radiation. In this paper, we have studied in an exact
way the emission of Hawking radiation both on the brane and in
the bulk from non-rotating, uncharged $(4+n)$-dimensional black holes,
and searched for distinctive features in the radiation spectra which would
allow us to determine the number of extra dimensions that exist in nature.
At the same time, we provided answers to questions that had remained open
from previous analyses in the literature. 

Focusing initially on the emission of brane-localized modes, we have derived
a {\it master equation} describing the motion of a field with arbitrary spin
$s$ in the spherically-symmetric black hole background induced on the brane.
We then provided exact, numerical results for the greybody factors and emission
rates for scalars, fermions and gauge bosons. These results are valid in
all energy regimes, and agree with the full analytic results derived in
\cite{kmr2} at low- and intermediate-energy scales. The greybody factors
interpolate between the low- and high-energy asymptotic limits in a way that
depends both on their spin and the dimensionality of spacetime: while they
adopt the same asymptotic values (which decrease as $n$ increases) for all 
particle species in the high-energy regime, in the low-energy limit the 
greybody factors are enhanced as $n$ increases for $s=1$ and $\frac{1}{2}$, and 
suppressed for $s=0$. Therefore, their implementation in the calculation
of the emission rates is imperative if one wants to derive accurate results
for these quantities. By doing that, the energy emission rates obtained
reveal a substantial enhancement as the number of extra dimensions projected
onto the brane increases. The enhancement amounts to orders of magnitude,
for large values of $n$ and for all particle species, although the
increase in the rate depends strongly on the spin of the particle studied.
The computed total relative emissivities reveal that scalar fields,
which are the dominant form of particles emitted by the black hole
for $n=0$, are outnumbered by the gauge bosons for large values of $n$,
with the fermions being the least effective channel during the emission.
Therefore, both the amount and the type of the emitted radiation by a
higher-dimensional black hole directly on the brane may possibly lead to
the determination of the number of extra dimensions.

The emission of particle modes on the brane is the most phenomenologically
interesting effect since it involves Standard Model particles that can be
easily detected during experiments. Nevertheless, a small higher-dimensional
black hole emits also bulk modes and inevitably a proportion of the total
energy is lost into the bulk. In the second part of this paper, we 
investigated the details of the emission of bulk scalar modes with the
final aim being to provide an accurate estimate for the relative
bulk-to-brane energy emissivity. The corresponding greybody factors
depend again on the dimensionality of spacetime and so do the
energy emission rates that exhibit a similar, although less substantial,
enhancement as $n$ increases. Comparing the total bulk and brane emission
rates for scalars, integrated over the whole energy regime, we conclude
that most of the energy of the black hole is emitted in the form of brane
modes for all values of $n$. The total emissivity in the bulk reduces
to less than 1/4 of that on the brane for $n=2,3,4$, while it becomes
substantial for the extreme value of $n=7$, without however exceeding
the brane value. The accurate results, produced here for the first time in the
literature, clarify the situation concerning the relative amounts of
energy emitted in the bulk and on the brane, and provide firm support
to the heuristic arguments made in Ref. \cite{emparan}.

We should note at this point that no results were presented in this paper
concerning the emission of gravitons either on the brane or in the bulk.
The derivation of a consistent equation, that can describe the motion
of gravitons on the induced black hole background on the brane, is still
under investigation. Nevertheless, we expect that, as in the case of the
emission of fields with spin $s=0,1$ and $\frac{1}{2}$, the graviton
emission becomes enhanced as the dimensionality of spacetime increases,
and that it remains subdominant compared to that of the other species at
least for small values of $n$. In the near future, we hope to be able to
present complete results for the emission rates for gravitons on the brane,
and for the relative bulk-to-brane graviton emissivity. 

A final comment needs to be made here concerning the validity of the results
derived in this work. As pointed out in the text, the horizon of the black hole
and its mass, or equivalently the energy needed for the production of the
miniature black hole, is an input parameter of the analysis. Our results are
applicable for all values of $r_H$ which are smaller than the size of the
extra dimensions $R$, no matter how small or large $R$ is (as long as it
remains considerably larger than $\ell_{Pl}$ to avoid quantum corrections). 
Our analysis therefore remains valid for all theories postulating the
existence of flat extra dimensions and a fundamental scale of gravity even a
few orders of magnitude lower than $M_{Pl}$. If the emission of Hawking
radiation from these black holes is successfully detected, either at
next-generation colliders or in the more distant future, the
distinctive features discussed here may help in the determination of
the number of extra dimensions that exist in nature.

\acknowledgments

We would like to thank John March-Russell for a constructive collaboration
at an early stage, and Bryan Webber for helpful discussions and suggestions
throughout this work.
P.K. would also like to thank V. Rubakov, M. Shaposhnikov and P. Tinyakov
for a stimulating discussion when this project had just begun, and
R. Emparan for useful comments on the manuscript. C.H. was funded during
this work by the U.K. Particle Physics and Astronomy Research Council.

\bigskip \bigskip  

\noindent {\large \bf Appendix}

\appendix

\bigskip \noindent
In this Appendix, we provide a few steps in the Newman-Penrose formalism for the
derivation of the {\it master equation} describing the motion of a particle with
spin $s$ in the background of a higher-dimensional,  non-rotating, neutral black
hole projected onto a 3-brane. The corresponding 4-dimensional metric tensor is
given in Eq. (\ref{non-rot}). We first need to choose a tetrad basis of null
vectors $(\ell^\mu, n^\mu, m^\mu, \bar m^\mu)$, where $\ell$ and $n$ are real
vectors and $m$ and $\bar m$ are a pair of complex conjugate vectors. They
satisfy the relations: ${\bf l} \cdot {\bf n}=1$, ${\bf m}\cdot {\bf \bar m}=-1$,
with all other products being zero. Such a tetrad basis is given by:
%%%%%%%%%%%
\begin{eqnarray}
\ell^\mu=\Bigl(\frac{1}{h},\,1, \,0,\,0\Bigr)\,, &\quad &
n^\mu=\Bigl(\frac{1}{2}, \,-\frac{h}{2}, \,0, \,0\Bigr)\,, \nonumber \\[3mm]
m^\mu=\Bigl(0, \,0, \,1, \,\frac{i}{\sin\theta}\Bigr)\,
\frac{1}{\sqrt{2} r}\,, &\quad&
\bar m^\mu=\Bigl(0, \,0, \,1, \,\frac{-i}{\sin\theta}\Bigr)\,
\frac{1}{\sqrt{2} r}\,. 
\end{eqnarray}
%%%%%%%%%%%

The $\lambda_{abc}$ coefficients, which are used to construct the {\it spin
coefficients}, are defined as:
$\lambda_{abc}=(e_b)_{i,j}\Bigl[(e_a)^i (e_c)^j-(e_a)^j (e_c)^i \Bigr]$,
where $e_a$ stands for each one of the null vectors and $(i,j)$ denote the
components of each vector. Their non-vanishing components are found to be:
%%%%%%%%%%%%%
\begin{equation}
\lambda_{122}=-\frac{h'}{2}\,, \quad \lambda_{134}=\frac{1}{r}\,,
\quad \lambda_{234}=-\frac{h}{2r}\,, \quad
\lambda_{334}=\frac{\cos\theta}{\sqrt{2} r \sin\theta}\,.
\end{equation}
%%%%%%%%%%%%%
The above components must be supplemented by those that follow from the
symmetry $\lambda_{abc}=-\lambda_{cba}$ and the complex conjugates obtained
by replacing an index 3 by 4 (or vice versa) or interchanging 3 and 4 (when 
they are both present).

We may now compute the {\it spin coefficients} defined by
$\gamma_{abc}=(\lambda_{abc}+\lambda_{cab} -\lambda_{bca})/2$. Particular
components, or combinations, of the spin coefficients can be directly used
in the field equations \cite{np, Chandra}. They are found to have the
values:
%%%%%%%%%%%%%
\begin{eqnarray}
\kappa=\sigma=\lambda=\nu=\tau=\pi=\epsilon=0\,, \nonumber
\end{eqnarray}
%%%
\begin{equation}
\rho=-\frac{1}{r}\,, \quad \mu=-\frac{h}{2r}\,,\quad \gamma=\frac{h'}{4}\,, 
\quad\alpha=-\beta=-\frac{\cot\theta}{2\sqrt{2} r}\,.
\end{equation}
%%%%%%%%%%%%

In what follows, we will also employ the Newman-Penrose operators:
%%%%%%%%%%%%%
\begin{equation}
\hat D=\frac{1}{h}\,\frac{\partial \,}{\partial t} +
\frac{\partial \,}{\partial r}\,, \qquad
\hat \Delta = \frac{1}{2}\,\frac{\partial \,}{\partial t} -
\frac{h}{2}\,\frac{\partial \,}{\partial r}\,, \qquad
\hat \delta = \frac{1}{\sqrt{2} r}\,\Bigl(\frac{\partial \,}{\partial \theta} +
\frac{i}{\sin\theta}\,\frac{\partial \,}{\partial \varphi}\Bigr)\,,
\end{equation}
%%%%%%%%%%%
and make use of the following field factorization:
\begin{equation}
\Psi_s(t,r,\theta,\varphi)= e^{-i\om t}\,e^{i m \varphi}\,R_{s}(r)
\,S^{m}_{s,\ell}(\theta)\,,
\label{facto}
\end{equation}
where $Y^m_{s,\ell}=e^{i m \varphi}\,S^{m}_{s,\ell}(\theta)$ are the spin-weighted
spherical harmonics \cite{goldberg}. We will now consider each type of field
separately:

\bigskip \medskip \noindent
\underline{{\bf I. Gauge Bosons ($s=1$)}}

\bigskip
In the Newman-Penrose formalism, there are only three `degrees of freedom'
for a gauge field, namely $\Phi_0=F_{13}$, $\Phi_1=(F_{12}+ F_{43})/2$ and
$\Phi_2=F_{42}$, in terms of which the different components of the Yang-Mills 
equation for a massless gauge field are written as:
%%%%%%%%%%%%%
\begin{eqnarray}
(\hat D-2 \rho)\,\Phi_1 - (\hat \delta^*-2 \alpha) \,\Phi_0 &=& 0\,, 
\label{B1}\\[3mm]
\hat \delta\,\Phi_1 - (\hat \Delta + \mu-2 \gamma) \,\Phi_0 &=& 0\,, 
\label{B2}\\[3mm]
(\hat D-\rho)\,\Phi_2 - \hat \delta^*\,\Phi_1 &=& 0\,, \label{B3}\\[3mm]
(\hat \delta+2 \beta)\,\Phi_2 - (\hat \Delta +2 \mu) \,\Phi_1 &=& 0\,, 
\label{B4}
\end{eqnarray}
%%%%%%%%%%%
where $\hat \delta^*$ stands for the complex conjugate of $\hat \delta$.
Rearranging Eqs. (\ref{B1})-(\ref{B2}), one can see that $\Phi_1$ decouples
leaving behind an equation involving only $\Phi_0$. Using the explicit forms
of the operators and spin coefficients, as well as the factorized ansatz
(\ref{facto}), this can be separated into an angular equation,
%%%%%%%%%%%%%
\begin{equation}
\frac{1}{\sin\theta}\,\frac{d \,}{d \theta}\,
\biggl(\sin\theta\,\frac{d S^m_{1,\ell}}{d \theta}\biggr)
+ \biggl[ -\frac{2 m \cot\theta}{\sin\theta} -\frac{m^2}{\sin^2\theta}
+ 1 - \cot^2\theta + \lambda_{1\ell} \biggl] S^m_{1,\ell}(\theta)=0\,,
\end{equation}
%%%%%%%%%%%%
with eigenvalue $\lambda_{s\ell}=\ell\,(\ell +1) - s\,(s+1)$, and a radial
equation:
%%%%%%%%%%%%%
\begin{equation}
\frac{1}{\Delta}\,\frac{d \,}{d r}\biggl(
\Delta^2\,\frac{d R_1}{d r}\biggr) + \biggl[
\frac{\omega^2 r^2}{h} + 2 i \omega r -\frac{i \omega r^2 h'}{h}
+ (\Delta'' -2) -\lambda_{1\ell}\biggr]\,R_1(r)=0\,,
\end{equation}
%%%%%%%%%%%%%
where we have defined $\Delta=h r^2$.

\bigskip \medskip \noindent
{\bf \underline{II. Fermion Fields ($s=1/2$)}}

\bigskip
For a massless two-component spinor field, the Dirac equation can be
written as:
%%%%%%%%%%%%
\begin{eqnarray}
(\hat \delta^*-\alpha)\,\chi_0 &=& (\hat D-\rho)\,\chi_1\,, \\[3mm]
(\hat \Delta + \mu -\gamma)\,\chi_0 &=& (\hat \delta + \beta)\,\chi_1\,.
\end{eqnarray}
%%%%%%%%%%%
Performing a similar rearrangement as in the case of bosons, we find that
$\chi_1$ is decoupled and that the equation for $\chi_0$ reduces to the
following set of angular,
%%%%%%%%%%%%%
\begin{equation}
\frac{1}{\sin\theta}\,\frac{d \,}{d \theta}\,
\biggl(\sin\theta\,\frac{d S^m_{1/2,\ell}}{d \theta}\biggr)
+ \biggl[ -\frac{m \cot\theta}{\sin\theta} -\frac{m^2}{\sin^2\theta}
+ \frac{1}{2} - \frac{1}{4}\,\cot^2\theta + \lambda_{\frac{1}{2}\ell} \biggl] 
S^m_{1/2,\ell}(\theta)=0\,,
\end{equation}
%%%%%%%%%%%%
and radial,
%%%%%%%%%%%%%
\begin{equation}
\frac{1}{\sqrt{\Delta}}\,\frac{d \,}{d r}\biggl(
\Delta^{3/2}\,\frac{d R_{1/2}}{d r}\biggr) + \biggl[
\frac{\omega^2 r^2}{h} + i \omega r -\frac{i \omega r^2 h'}{2h}
+ \frac{1}{2}\,(\Delta'' -2) -\lambda_{\frac{1}{2}\ell}\biggr]\,R_{1/2}(r)=0\,,
\end{equation}
%%%%%%%%%%%%%
equations, with the same definitions for $\Delta$ and $\lambda_{s\ell}$
as before.

\bigskip \medskip \noindent
{\bf \underline{III. Scalar Fields ($s=0$)}}

\bigskip
For completeness, we add here the equation of motion for a scalar field
propagating in the same background. This equation can be determined quite
easily by evaluating the double covariant derivative $g^{\mu\nu} D_\mu D_\nu$
acting on the scalar field. It finally leads to the pair of equations
%%%%%%%%
\begin{equation}
\frac{1}{\sin\theta}\,\frac{d \,}{d \theta}\,\biggl(\sin\theta\,
\frac{d S^m_{0,\ell}}{d \theta}\,\biggr) + \biggl[-\frac{m^2}{\sin^2\theta}
+ \lambda_{0\ell} \biggr]\,S^m_{0,\ell}=0\,,
\end{equation}
%%%%%%%%%%%
%%%%%%%%%%%%
\begin{equation}
\frac{d \,}{dr}\,\biggl(\Delta\,\frac{d R_0}{dr}\biggr) +
\Bigl(\frac{\om^2 r^2}{h} - \lambda_{0\ell}\Bigr) R_0(r) =0\,,
\label{scalar}
\end{equation}
%%%%%%%%%%%
where $Y^m_{0,\ell}=e^{i m \varphi}\,S^{m}_{0,\ell}(\theta)$ are the usual
spherical harmonics $Y^m_\ell(\theta, \varphi)$ and
$\lambda_{0\ell}=\ell (\ell+1)$.
The above equations were used in \cite{kmr1} for the analytic determination
of the greybody factors for the emission of scalar particles on the brane by a
higher-dimensional black hole.

\bigskip \medskip \noindent
{\bf \underline{IV. Master Equation for a field with arbitrary spin $s$}}

\bigskip
Combining all the above equations derived for bosons, fermions
and scalar fields, we may now rewrite them in the form of a {\it master
equation} valid for all types of fields. The radial equation then takes
the form:
\begin{equation}
\Delta^{-s}\,\frac{d \,}{dr}\,\biggl(\Delta^{s+1}\,\frac{d R_s}{dr}\,\biggr) +
\biggl(\frac{\om^2 r^2}{h} + 2i s\,\om\,r -\frac{i s \om\,r^2 h'}{h}
+s\,(\Delta''-2) - \lambda_{s\ell} \biggr)\,R_s(r)=0\,,
\label{master1}
\end{equation}
%%%%%%%%%%%
while, the angular equation reads
%%%%%%%%
\begin{equation}
\frac{1}{\sin\theta}\,\frac{d \,}{d \theta}\,\biggl(\sin\theta\,
\frac{d S^m_{s,\ell}}{d \theta}\,\biggr) + \biggl[-\frac{2 m s \cot\theta}
{\sin\theta} - \frac{m^2}{\sin^2\theta} + s - s^2 \cot^2\theta 
+ \lambda_{s\ell} \biggr]\,S^m_{s,\ell}=0\,.
\label{master2}
\end{equation}
%%%%%%%%%%%
The latter equation is identical to the one derived by Teukolsky \cite{Teukolsky}
in the case of a non-rotating, spherically-symmetric black hole. The radial one
differs by the extra factor $s\,(\Delta''-2)$ due to the fact that for our metric
tensor this combination is not zero, contrary to what happens in the case of the
4-dimensional Schwarzschild, or Kerr, metric. The $\Delta''$- term can be removed
if we make the redefinition $R_s=\Delta^{-s}\, P_s$.  Then, we obtain:
%%%%%%%%%%%%%%
\begin{equation}
\Delta^{s}\,\frac{d \,}{dr}\,\biggl(\Delta^{1-s}\,\frac{d P_s}{dr}\,\biggr) +
\biggl(\frac{\om^2 r^2}{h} + 2i s\,\om\,r -\frac{i s \om\,r^2 h'}{h}
- \Lambda \biggr)\,P_s(r)=0\,,
\label{master3}
\end{equation}
%%%%%%%%%%%
where now $\Lambda=\lambda_{s\ell} +2 s =\ell\,(\ell+1)-s\,(s-1)$. The above
form of the radial equation was used in \cite{kmr2} to determine in an 
analytic way the greybody factors and
emission rates for the emission of fermions and gauge bosons on the brane
by a higher-dimensional, spherically-symmetric black hole.


\begin{thebibliography}{99}
\bibitem{ADD}
N.~Arkani-Hamed, S.~Dimopoulos and G.~Dvali,
{\it The Hierarchy Problem and New Dimensions at a Millimeter},
\plb{429}{1998}{263} [\hepph{9803315}];\\[1mm]
%%CITATION = HEP-PH 9803315;%%
N.~Arkani-Hamed, S.~Dimopoulos and G.~Dvali,
{\it Phenomenology, astrophysics and cosmology of theories with
sub-millimeter dimensions and TeV scale quantum gravity},
\prd{59}{1999}{086004} [\hepph{9807344}];\\[1mm]
%%CITATION = HEP-PH 9807344;%%
I.~Antoniadis, N.~Arkani-Hamed, S.~Dimopoulos and G.~R.~Dvali,
{\it New dimensions at a millimeter to a Fermi and superstrings at a TeV},
\plb{436}{1998}{257} [\hepph{9804398}].
%%CITATION = HEP-PH 9804398;%%

\bibitem{RS}
L.~Randall and R.~Sundrum,
{\it A Large Mass Hierarchy from a Small Extra Dimension},
\prl{83}{1999}{3370} [\hepph{9905221}];\\[1mm]
%%CITATION = HEP-PH 9905221;%%
L.~Randall and R.~Sundrum, {\it An alternative to compactification},
\prl{\bf 83}{1999}{4690} [\hepth{9906064}].
%%CITATION = HEP-TH 9906064;%%

\bibitem{early}
K.~Akama, {\it An Early Proposal Of 'Brane World'},
{\it Lect.\ Notes Phys.}\  {\bf 176} (1982) 267 [\hepth{0001113}];\\[1mm]
%%CITATION = HEP-TH 0001113;%%
%%
V.~A.~Rubakov and M.~E.~Shaposhnikov,
{\it Extra spacetime Dimensions: Towards A Solution To The Cosmological
Constant Problem},
\plb{125}{1983}{139}; \\[1mm] %%CITATION = PHLTA,B125,139;%%
%%
V.~A.~Rubakov and M.~E.~Shaposhnikov, {\it Do We Live Inside A Domain Wall?},
\plb{125}{1983}{136};\\[1mm] %%CITATION = PHLTA,B125,136;%%
%%
M.~Visser, {\it An Exotic Class Of Kaluza-Klein Models},
\plb{159}{1985}{22} [\hepth{9910093}];\\[1mm] %%CITATION = HEP-TH 9910093;%%
%%
I.~Antoniadis, {\it A Possible New Dimension At A Few Tev},
\plb{246}{1990}{377};\\[1mm] %%CITATION = PHLTA,B246,377;%%
%%
I.~Antoniadis, K.~Benakli and M.~Quiros,
{\it Production of Kaluza-Klein states at future colliders},
\plb{331}{1994}{313} [\hepph{9403290}];\\[1mm]
%%CITATION = HEP-PH 9403290;%%
%%
J.~Lykken, {\it Weak Scale Superstrings},
\prd{54}{1996}{3693} [\hepth{9603133}]. %%CITATION = HEP-TH 9603133;%%

\bibitem{Hoyle}
C.~D.~Hoyle, U.~Schmidt, B.~R.~Heckel, E.~G.~Adelberger, J.~H.~Gundlach,
D.~J.~Kapner and H.~E.~Swanson,
{\it Sub-millimeter tests of the gravitational inverse-square law: A search 
for 'large' extra dimensions},
\prl{86}{2001}{1418} [\hepph{0011014}]. %%CITATION = HEP-PH 0011014;%%

\bibitem{colliders}
P.~Abreu {\it et al.}  [DELPHI Collaboration],
{\it Photon Events with Missing Energy at $\sqrt{s} = 183$ to 189 GeV},
\epjc{17}{2000}{53} [\hepex{0103044}];\\[1mm] %%CITATION = HEP-EX 0103044;%%
%
G.~Abbiendi {\it et al.}  [OPAL Collaboration],
{\it Photonic events with missing energy in e+ e- collisions at 
s**(1/2) =  189-GeV},''
\epjc{18}{2000}{253} [\hepex{0005002}];\\[1mm] %%CITATION = HEP-EX 0005002;%%
%
D.~Acosta {\it et al.}  [CDF Collaboration],
{\it Limits on extra dimensions and new particle production in the exclusive
photon and missing energy signature in p anti-p collisions at
s**(1/2)  = 1.8-TeV},
\prl{89}{2002}{281801} [\hepex{0205057}]. %%CITATION = HEP-EX 0205057;%%

\bibitem{astro}
S.~Cullen and M.~Perelstein,
{\it SN1987A constraints on large compact dimensions},
\prl{83}{1999}{268} [\hepph{9903422}]; \\[1mm] 
%%CITATION = HEP-PH 9903422;%%
L.~J.~Hall and D.~R.~Smith,
{\it Cosmological constraints on theories with large extra dimensions},
\prd{60}{1999}{085008} [\hepph{9904267}];\\[1mm]
%%CITATION = HEP-PH 9904267;%%
V.~D.~Barger, T.~Han, C.~Kao and R.~J.~Zhang,
{\it Astrophysical constraints on large extra dimensions},
\plb{461}{1999}{34} [\hepph{9905474}];\\[1mm] %%CITATION = HEP-PH 9905474;%%
C.~Hanhart, D.~Phillips, S.~Reddy and M.~Savage,
{\it Extra dimensions, SN1987a, and nucleon nucleon scattering data},
\npb{595}{2001}{335} [\nuclth{0007016}];\\[1mm] 
%%CITATION = NUCL-TH 0007016;%%
C.~Hanhart, J.~A.~Pons, D.~R.~Phillips and S.~Reddy,
{\it The likelihood of GODs' existence: Improving the SN 1987a constraint on 
the size of large compact dimensions},
\plb{509}{2001}{1} [\astroph{0102063}];\\[1mm]
%%CITATION = ASTRO-PH 0102063;%%
S.~Hannestad,
{\it Strong constraint on large extra dimensions from cosmology},
\prd{64}{2001}{023515} [\hepph{0102290}];\\[1mm]
%%CITATION = HEP-PH 0102290;%%
S.~Hannestad and G.~Raffelt,
{\it New supernova limit on large extra dimensions},
\prl{87}{2001}{051301} [\hepph{0103201}];\\[1mm]
%%CITATION = HEP-PH 0103201;%%
S.~Hannestad and G.~G.~Raffelt,
{\it Stringent neutron-star limits on large extra dimensions},
\prl{88}{2002}{071301} [\hepph{0110067}];\\[1mm]
%%CITATION = HEP-PH 0110067;%%
R.~Allahverdi, C.~Bird, S.~Groot Nibbelink and M.~Pospelov,
{\it Cosmological bounds on large extra dimensions from non-thermal
production of Kaluza-Klein modes}, \hepph{0305010};\\[1mm]
%%CITATION = HEP-PH 0305010;%%
L.~A.~Anchordoqui, J.~L.~Feng, H.~Goldberg and A.~D.~Shapere,
{\it Updated limits on TeV-scale gravity from absence of neutrino cosmic
ray showers mediated by black holes}, \hepph{0307228}.
%%CITATION = HEP-PH 0307228;%%

\bibitem{ADMR}
P.~C.~Argyres, S.~Dimopoulos and J.~March-Russell,
{\it Black holes and sub-millimeter dimensions},
\plb{441}{1998}{96} [\hepth{9808138}]. %%CITATION = HEP-TH 9808138;%%

\bibitem{BF}
T.~Banks and W.~Fischler,
%``A model for high-energy scattering in quantum gravity,''
\hepth{9906038}. %%CITATION = HEP-TH 9906038;%%

\bibitem{GT}
S.~B.~Giddings and S.~Thomas,
{\it High Energy Colliders as Black Hole Factories: The End of Short
Distance Physics}, \prd{65}{2002}{056010} [\hepph{0106219}].
%%CITATION = HEP-PH 0106219;%%

\bibitem{dl}
S.~Dimopoulos and G.~Landsberg,
{\it Black holes at the LHC},
\prl{87}{2001}{161602} [\hepph{0106295}].%%CITATION = HEP-PH 0106295;%%

\bibitem{voloshin}
M.~B.~Voloshin, {\it Semiclassical suppression of black hole production
in particle collisions}, \plb{518}{2001}{137} [\hepph{0107119}];\\[1mm]
%%CITATION = HEP-PH 0107119;%%
M.~B.~Voloshin, {\it More Remarks on Suppression of Large Black Hole
Production in Particle Collisions}, \plb{524}{2002}{376} [\hepph{0111099}].
%%CITATION = HEP-PH 0111099;%%

\bibitem{DE}
S.~Dimopoulos and R.~Emparan,
{\it String balls at the LHC and beyond},
\plb{526}{2002}{393} [\hepph{0108060}].%%CITATION = HEP-PH 0108060;%%

\bibitem{giddings}
S.~B.~Giddings,
{\it Black Hole Production in TeV-scale Gravity, and the Future of High
Energy Physics}, \hepph{0110127}. %%CITATION = HEP-PH 0110127;%%

\bibitem{GRW}
G.~F.~Giudice, R.~Rattazzi and J.~D.~Wells,
{\it Transplanckian collisions at the LHC and beyond},
\npb{630}{2002}{293} [\hepph{0112161}]. %%CITATION = HEP-PH 0112161;%%

\bibitem{Thorne}
K.~S.~Thorne, {\it Nonspherical Gravitational Collapse: A Short Review},
in {\it Magic without Magic}, Ed. J.~R.~Klauder (San Fransisco, 1972).

\bibitem{fraction}
R. Penrose, {\it unpublished}, 1974;\\[1mm]
P.~D.~D'Eath and P.~N.~Payne,
{\it Gravitational Radiation In High Speed Black Hole Collisions. 1.
Perturbation Treatment Of The Axisymmetric Speed Of Light Collision},
\prd{46}{1992}{658};\\[1mm] %%CITATION = PHRVA,D46,658;%%
D.~M.~Eardley and S.~B.~Giddings,
{\it Classical Black Hole Production in High-Energy Collisions}, 
\prd{66}{2002}{044011} [\grqc{0201034}];\\[1mm]
%%CITATION = GR-QC 0201034;%%
H.~Yoshino and Y.~Nambu,
{\it High-energy head-on collisions of particles and hoop conjecture},
\prd{66}{2002}{065004} [\grqc{0204060}]. %%CITATION = GR-QC 0204060;%%


\bibitem{cosmic1}
A.~Goyal, A.~Gupta and N.~Mahajan,
{\it Neutrinos as source of ultra high-energy cosmic rays in extra dimensions},
\prd{63}{2001}{043003} [\hepph{0005030}];\\[1mm]
%%CITATION = HEP-PH 0005030;%%
J.~L.~Feng and A.~D.~Shapere, {\it Black hole production by cosmic rays},
\prl{88}{2002}{021303} [\hepph{0109106}];\\[1mm]
%%CITATION = HEP-PH 0109106;%%
%%
L.~Anchordoqui and H.~Goldberg,
{\it Experimental signature for black hole production in neutrino air showers},
\prd{65}{2002}{047502} [\hepph{0109242}];\\[1mm]
%%CITATION = HEP-PH 0109242;%%
%%
R.~Emparan, M.~Masip and R.~Rattazzi,
{\it Cosmic rays as probes of large extra dimensions and TeV gravity},
\prd{65}{2002}{064023} [\hepph{0109287}];\\[1mm]
%%CITATION = HEP-PH 0109287;%%
L.~A.~Anchordoqui, J.~L.~Feng, H.~Goldberg and A.~D.~Shapere,
{\it Black holes from cosmic rays: Probes of extra dimensions and new limits
on TeV-scale gravity},
\prd{65}{2002}{124027} [\hepph{0112247}];\\[1mm]
%%CITATION = HEP-PH 0112247;%%
Y.~Uehara, {\it Production and detection of black holes at neutrino array},
\ptp{107}{2002}{621} [\hepph{0110382}];\\[1mm]
%%CITATION = HEP-PH 0110382;%%
J.~Alvarez-Muniz, J.~L.~Feng, F.~Halzen, T.~Han and D.~Hooper,
{\it Detecting microscopic black holes with neutrino telescopes},
\prd{65}{2002}{124015} [\hepph{0202081}].
%%CITATION = HEP-PH 0202081;%%

\bibitem{cosmic2}
A.~Ringwald and H.~Tu,
{\it Collider versus cosmic ray sensitivity to black hole production},
\plb{525}{2002}{135} [\hepph{0111042}];\\[1mm]
%%CITATION = HEP-PH 0111042;%%
M.~Kowalski, A.~Ringwald and H.~Tu, {\it Black holes at neutrino telescopes},
\plb{529}{2002}{1} [\hepph{0201139}];\\[1mm] %%CITATION = HEP-PH 0201139;%%
D.~Kazanas and A.~Nicolaidis, {\it Cosmic rays and large extra dimensions},
{\it Gen.\ Rel.\ Grav.} {\bf 35} (2003) 1117 [\hepph{0109247}];\\[1mm]
%%CITATION = HEP-PH 0109247;%%
P.~Jain, S.~Kar, S.~Panda and J.~P.~Ralston,
{\it Brane-production and the neutrino nucleon cross section at ultra high
energies in low scale gravity models}, \hepph{0201232};\\[1mm]
%%CITATION = HEP-PH 0201232;%%
A.~Ringwald, {\it Production of black holes in TeV-scale gravity},
{\it Fortsch.\ Phys.} {\bf 51} (2003) 830 [\hepph{0212342}];\\[1mm]
%%CITATION = HEP-PH 0212342;%%
E.~J.~Ahn, M.~Ave, M.~Cavaglia and A.~V.~Olinto,
{\it TeV black hole fragmentation and detectability in extensive air-showers},
\prd{68}{2003}{043004} [\hepph{0306008}];\\[1mm]
%%CITATION = HEP-PH 0306008;%%
A.~Nicolaidis and N.~G.~Sanchez,
{\it Signatures of TeV scale gravity in high energy collisions},
\hepph{0307321};\\[1mm] %%CITATION = HEP-PH 0307321;%%
L.~A.~Anchordoqui, J.~L.~Feng, H.~Goldberg and A.~D.~Shapere,
{\it Updated limits on TeV-scale gravity from absence of neutrino cosmic ray
showers mediated by black holes}, \hepph{0307228}.
%%CITATION = HEP-PH 0307228;%%


\bibitem{bhphen1}
S.~Hossenfelder, S.~Hofmann, M.~Bleicher and H.~Stocker,
{\it Quasi-stable black holes at LHC},
\prd{66}{101502}{2002} [\hepph{0109085}];\\[1mm]
%%CITATION = HEP-PH 0109085;%%
H.~C.~Kim, S.~H.~Moon and J.~H.~Yee,
%``Dimensional dependence of black hole formation in scalar field collapse,''
\jhep{0202}{2002}{046} [\grqc{0108071}];\\[1mm]
%%CITATION = GR-QC 0108071;%%
K.~Cheung, {\it Black hole production and large extra dimensions},
\prl{88}{2002}{221602} [\hepph{0110163}]; \\[1mm]
%%CITATION = HEP-PH 0110163;%%
S.~Hossenfelder, S.~Hofmann, M.~Bleicher and H.~Stocker,
{\it Black hole production in large extra dimensions at the Tevatron: A 
chance to observe a first glimpse of TeV scale gravity},
\plb{548}{200}{73} [\hepph{0112186}];\\[1mm]
%%CITATION = HEP-PH 0112186;%%
R.~Casadio and B.~Harms,
{\it Can black holes and naked singularities be detected in accelerators?},
\ijmpa{17}{2002}{4635} [\hepth{0110255}];\\[1mm]
%%CITATION = HEP-TH 0110255;%%
S.~C.~Park and H.~S.~Song,
{\it Production of spinning black holes at colliders},
{\it J.\ Korean Phys.\ Soc.}  {\bf 43} (2003) 30 [\hepph{0111069}];\\[1mm]
%%CITATION = HEP-PH 0111069;%%
G.~Landsberg, {\it Discovering new physics in the decays of black holes},
\prl{88}{2002}{181801} [\hepph{0112061}].
%%CITATION = HEP-PH 0112061;%%

\bibitem{bhphen2}
E.~J.~Ahn, M.~Cavaglia and A.~V.~Olinto, {\it Brane factories},
\plb{551}{2003}{1} [\hepth{0201042}];\\[1mm]
%%CITATION = HEP-TH 0201042;%%
T.~G.~Rizzo,
{\it Black hole production at the LHC: Effects of Voloshin suppression},
\jhep{0202}{2002}{011} [\hepph{0201228}];\\[1mm]
%%CITATION = HEP-PH 0201228;%%
S.~N.~Solodukhin,
{\it Classical and quantum cross-section for black hole production in 
particle collisions},
\plb{533}{2002}{153} [\hepph{0201248}];\\[1mm]
%%CITATION = HEP-PH 0201248;%%
V.~Cardoso and J.~P.~Lemos,
{\it Gravitational radiation from collisions at the speed of light: A 
massless particle falling into a Schwarzschild black hole},
\plb{538}{2002}{1} [\grqc{0202019}];\\[1mm] %%CITATION = GR-QC 0202019;%%
K.~Cheung, {\it Black hole, string ball, and p-brane production at hadronic
supercolliders},
\prd{66}{2002}{036007} [\hepph{0205033}];\\[1mm]
%%CITATION = HEP-PH 0205033;%%
R.~A.~Konoplya,
{\it On quasinormal modes of small Schwarzschild-anti-de-Sitter black hole},
\prd{\bf 66}{2002}{044009} [\hepth{0205142}];\\[1mm]
%%CITATION = HEP-TH 0205142;%%
Y.~Uehara, {\it New potential of black holes: Quest for TeV-scale physics
by measuring  top quark sector using black holes},
\hepph{0205199};\\[1mm]
%%CITATION = HEP-PH 0205199;%%
R.~Guedens, D.~Clancy and A.~R.~Liddle,
{\it Primordial black holes in braneworld cosmologies. I: Formation,
cosmological evolution and evaporation},
\prd{66}{2002}{043513} [\astroph{0205149}].
%%CITATION = ASTRO-PH 0205149;%%

\bibitem{bhphen3}
A.~V.~Kotwal and C.~Hays,
{\it Production and decay of spinning black holes at colliders and tests of
black hole dynamics}, \prd{66}{2002}{116005} [\hepph{0206055}];\\[1mm]
%%CITATION = HEP-PH 0206055;%%
V.~Frolov and D.~Stojkovic,
{\it Black hole radiation in the brane world and recoil effect},
\prd{66}{2002}{084002} [\hepth{0206046}];\\[1mm]
%%CITATION = HEP-TH 0206046;%%
A.~Chamblin and G.~C.~Nayak,
{\it Black hole production at LHC: String balls and black holes from pp
and  lead lead collisions},
\prd{66}{2002}{091901} [\hepph{0206060}];\\[1mm] 
%%CITATION = HEP-PH 0206060;%%
T.~Han, G.~D.~Kribs and B.~McElrath,
{ \it Black hole evaporation with separated fermions},
\prl{90}{2003}{031601} [\hepph{0207003}];\\[1mm]
%%CITATION = HEP-PH 0207003;%%
V.~Frolov and D.~Stojkovic,
{\it Black hole as a point radiator and recoil effect on the brane world},
\prl{89}{2002}{151302} [\hepth{0208102}];\\[1mm]
%%CITATION = HEP-TH 0208102;%%
L.~Anchordoqui and H.~Goldberg, {\it Black hole chromosphere at the LHC},
\prd{67}{2003}{064010} [\hepph{0209337}].
%%CITATION = HEP-PH 0209337;%%

\bibitem{bhphen4}
V.~Frolov and D.~Stojkovic,
{\it Quantum radiation from a 5-dimensional rotating black hole},
\prd{67}{2003}{084004} [\grqc{0211055}];\\[1mm]
%%CITATION = GR-QC 0211055;%%
V.~Cardoso and J.~P.~Lemos,
{\it Gravitational radiation from the radial infall of highly relativistic
point particles into Kerr black holes},
\prd{67}{2003}{084005} [\grqc{0211094}];\\[1mm]
%%CITATION = GR-QC 0211094;%%
V.~Cardoso, O.~J.~Dias and J.~P.~Lemos,
{\it Gravitational radiation in D-dimensional spacetimes},
\prd{67}{2003}{064026} [\hepth{0212168}];\\[1mm]
%%CITATION = HEP-TH 0212168;%%
V.~Frolov and D.~Stojkovic,
{\it Particle and light motion in a space-time of a five-dimensional
rotating black hole}, \grqc{0301016};\\[1mm]
%%CITATION = GR-QC 0301016;%%
A.~Chamblin, F.~Cooper and G.~C.~Nayak,
{\it Interaction of a TeV scale black hole with the quark gluon plasma at 
LHC}, \hepph{0301239};\\[1mm] %%CITATION = HEP-PH 0301239;%%
I.~Mocioiu, Y.~Nara and I.~Sarcevic,
{\it Hadrons as signature of black hole production at the LHC},
\plb{557}{2003}{87} [\hepph{0301073}];\\[1mm]
%%CITATION = HEP-PH 0301073;%%
R.~A.~Konoplya,
{\it Quasinormal behavior of the d-dimensional Schwarzshild black hole
and  higher order WKB approach},
\prd{\bf 68} {2003}{024018} [\grqc{0303052}];\\
%%CITATION = GR-QC 0303052;%%
R.~Casadio,
{\it On brane-world black holes and short scale physics},
\hepph{0304099};\\[1mm] %%CITATION = HEP-PH 0304099;%%
O.~Vasilenko,
{\it Trap surface formation in high-energy black holes collision},
\hepth{03\-05\-067};\\[1mm] %%CITATION = HEP-TH 0305067;%%
M.~Cavaglia, S.~Das and R.~Maartens,
{\it Will we observe black holes at LHC?},
\cqg{20}{2003}{L205} [\hepph{0305223}];\\
%%CITATION = HEP-PH 0305223;%%
R.~A.~Konoplya,
{\it Gravitational quasinormal radiation of higher-dimensional black holes},
\hepth{0309030}. %%CITATION = HEP-TH 0309030;%%


\bibitem{hawking}
S.~W.~Hawking, 
{\it Particle Creation by Black Holes}, \cmp{43}{1975}{199}.
%%CITATION = CMPHA,43,199;%%

\bibitem{page}
D.~N.~Page
{\it Particle Emission Rates from a Black Hole: Massless Particles from
an Uncharged, Nonrotating Hole}, \prd{13}{1976}{198}.
%%CITATION = PHRVA,D13,198;%%

\bibitem{kmr1}
P.~Kanti and J.~March-Russell,
{\it Calculable Corrections to Brane Black Hole Decay. 1.~The Scalar Case},
\prd{66}{2002}{024023} [\hepph{0203223}]. %%CITATION = HEP-PH 0203223;%%

\bibitem{kmr2}
P.~Kanti and J.~March-Russell,
{\it Calculable Corrections to Brane Black Hole Decay. 2.~Greybody Factors
for Spin 1/2 and 1},
\prd{67}{2003}{104019} [\hepph{0212199}]. %%CITATION = HEP-PH 0212199;%%

\bibitem{emparan}
R.~Emparan, G.~T.~Horowitz and R.~C.~Myers,
{\it Black Holes Radiate Mainly on the Brane}, \prl{85}{2000}{499}
[\hepth{0003118}]. %%CITATION = HEP-TH 0003118;%%

\bibitem{myers}
R.~C.~Myers and M.~J.~Perry,
{\it Black Holes in Higher Dimensional spacetimes}, \ap{172}{1986}{304}.
%%CITATION = APNYA,172,304;%%

\bibitem{GTK} S.~S.~Gubser, I.~R.~Klebanov and A.~A.~Tseytlin,
{\it String theory and classical absorption by three-branes},
\npb{499}{1997}{217} [\hepth{9703040}].
%%CITATION = HEP-TH 9703040;%%

\bibitem{HRW}
C.~M.~Harris, P.~Richardson and B.~R.~Webber,
{\it ``CHARYBDIS: A black hole event generator}, \jhep{0308}{2003}{033}
[\hepph{0307305}]. %%CITATION = HEP-PH 0307305;%% 

\bibitem{MTW}
C.~W.~Misner, K.~T.~Thorne and J.~A.~Wheeler, {\it Gravitation}
(Freeman, San Fransisco, 1973). 

\bibitem{sanchez}
N.~Sanchez, {\it Absorption And Emission Spectra Of A Schwarzschild Black
Hole}, \prd{18}{1978}{1030};\\[1mm] %%CITATION = PHRVA,D18,1030;%%
N.~Sanchez, {\it Elastic Scattering of Waves by a Black Hole},
\prd{18}{1978}{1798}. %%CITATION = PHRVA,D18,1798;%%


\bibitem{macG}
J.~H.~MacGibbon and B.~R.~Webber,
{\it Quark- and Gluon-jet Emission from Primordial Black Holes: The
Instantaneous Spectra}, \prd{41}{1990}{3052}.
%%CITATION = PHRVA,D41,3052;%%

\bibitem{Teukolsky}
S.~A.~Teukolsky, {\it Perturbations of a Rotating Black Hole. I. Fundamental
Equations for Gravitational, Electromagnetic and Neutrino-Field Perturbations},
{\it Astrophys. J.} {\bf 185} (1973) 635. %%CITATION = ASJOA,185,635;%%

\bibitem{np}
E.~Newman and R.~Penrose, {\it An approach to Gravitational Radiation by
a Method of Spin Coefficients}, \jmp{3}{1962}{566}. 
%%CITATION = JMAPA,3,566;%%

\bibitem{Chandra}
S.~Chandrasekhar, {\it The Mathematical Theory of Black Holes} (Oxford
University Press, New York, 1983).

\bibitem{goldberg} J.~N.~Goldberg, A.~J.~MacFarlane, E.~T.~Newman,
F.~Rohrlich and E.~C.~Sudarshan,
{\it Spin S Spherical Harmonics And Edth},
\jmp{8}{1967}{2155}. %%CITATION = JMAPA,8,2155;%%

\bibitem{CL} M.~Cvetic and F.~Larsen,
{\it Greybody factors for black holes in four dimensions: Particles with 
spin}, \prd{57}{1998}{6297} [\hepth{9712118}].
%%CITATION = HEP-TH 9712118;%%

\bibitem{iop}
D.~Ida, K-y.~Oda and S.~C.~Park,
{\it Rotating Black Holes at Future Colliders: Greybody Factors for Brane
Fields}, \prd{67}{2003}{064026} [\hepth{0212108}]. 
%%CITATION = HEP-TH 0212108;%%

%\bibitem{tp2}
%W.~H.~Press and S.~A.~Teukolsky,
%{\it Perturbations of a Rotating Black Hole. II.~Dynamical Stability of the
% Kerr Metric},
%{\it Astrophys. J.} {\bf 185} (1973) 649.

\bibitem{tp3}
W.~H.~Press and S.~A.~Teukolsky,
{\it Perturbations of a Rotating Black Hole. III.~Interaction of the Hole
with Gravitational and Electromagnetic Radiation},
{\it Astrophys. J.} {\bf 193} (1974) 443. %%CITATION = ASJOA,193,443;%%


\bibitem{cavaglia}
M.~Cavaglia,
{\it Black hole multiplicity at particle colliders (Do black holes radiate
mainly on the brane?)}, \hepph{0305256}.
%%CITATION = HEP-PH 0305256;%%

\end{thebibliography}
\end{document}